\documentclass{PoS}

\usepackage{amsmath}

\usepackage{slashed}

\usepackage{xspace}
\usepackage{cite}

\usepackage{cleveref}

\newcommand{\GeV}{{\ensuremath\rm GeV}}
\newcommand{\TeV}{{\ensuremath\rm TeV}}
\newcommand{\eqn}{equation}
\newcommand{\lam}{\lambda}
\newcommand{\pb}{\ensuremath\rm pb}
\newcommand{\fb}{\ensuremath\rm fb}
\newcommand{\ab}{\ensuremath\rm ab}
\newcommand{\al}{\alpha}
\newcommand{\be}{\beta}

\newcommand{\lb}{\left(}
\newcommand{\rb}{\right)}
\newcommand{\eqdot}{\,.}
\newcommand{\Ztwo}{\ensuremath{{\mathbb{Z}_2}}\xspace}
\newcommand{\eqcomma}{\,,}

\newcommand{\HiggsBounds}{\texttt{HiggsBounds}\xspace}
\newcommand{\HiggsSignals}{\texttt{HiggsSignals}\xspace}

\title{Investigating extended scalar sectors at current and future colliders}

\ShortTitle{Extended scalar sectors}

\author{\speaker{Tania Robens}\\
        Division of Theoretical Physics, Ruder Boskovic Institute, Zagreb, Croatia\\
        E-mail: \email{trobens@irb.hr}}

\abstract{In this work, I briefly report on constraints that can be obtained on new physics models that extend the scalar sector of the Standard Model (SM) of particle physics at the LHC. I concentrate on a few simple examples which serve to demonstrate advantages as well as possible drawbacks of current experimental searches, and comment on the discovery prospects of such models at future colliders. 
}

\FullConference{7th Annual Conference on Large Hadron Collider Physics - LHCP2019\\
		20-25 May, 2019\\
		Puebla, Mexico}

\begin{document}

\section{Introduction}
During Run-II of the Large Hadron Collider (LHC) at CERN, both the ATLAS and CMS experiments have collected more than $\int\mathcal{L}\,=\,150\,\fb^{-1}$ of data, with results using the full dataset starting to become available \cite{atlpub,cmspub}. One important quest for the theoretical and experimental community is the understanding of the scalar sector realized by nature. After the discovery of the Higgs boson in 2012 \cite{Aad:2012tfa,Chatrchyan:2012xdj}, many measurements have confirmed that the discovered boson complies with the properties of the Higgs boson predicted by the SM alone; however, both theoretical and experimental uncertainties still allow for beyond the SM (BSM) models with additional particle content. Understanding the constraints on such models, including detailed investigations of possible discovery channels, is an important task for the experimental and theoretical community.

There are many models that allow for such an extension of the SM scalar sector. In the following, I will focus on three examples that serve to identify strengths
 as well as possible drawbacks of current and future collider searches. I will first discuss the simplest extension, namely, the introduction of a real scalar which is a singlet under the SM gauge group \cite{Schabinger:2005ei,Patt:2006fw}. This model has a small number of free parameters, and its study can therefore serve to understand the possible interplay of theoretical and experimental constraints within a relatively simple framework. I will then turn to the Inert Doublet Model, a two Higgs doublet model featuring a dark matter (DM) candidate \cite{Deshpande:1977rw,Barbieri:2006dq,Cao:2007rm}. This model renders signatures that are already investigated by the LHC experiments, although currently no dedicated interpretation exists within an experimental analysis. I will briefly comment on possible alterations of the current search strategies which are needed in order to investigate this model at the LHC, as well as discovery prospects at future $e^+ e^-$ machines. Finally, I will turn to scalar-to-scalar decays that are currently not investigated by the experimental collaborations, and provide a short overview on interesting new signatures within a specific model, that are produced with feasible rates at the 13 \TeV~ LHC and require new experimental search strategies.\\

In the analysis of the models discussed here, various tools have been used for the determination of viable regions in the respective models parameter spaces. For the inclusion of signal strength measurements of the 125 \GeV~scalar as well as null-results for additional scalar searches, we made use of the publicly available tools \HiggsBounds\texttt{-5.4.0}~\cite{Bechtle:2008jh,Bechtle:2011sb,Bechtle:2013gu,Bechtle:2013wla, Bechtle:2015pma} and \HiggsSignals\texttt{-2.3.0}~\cite{Stal:2013hwa, Bechtle:2013xfa,Bechtle:2014ewa}.

\section{Simple scalar extension of the SM scalar sector}
The simplest extension of the SM scalar sector introduces an additional real scalar which is a singlet under the SM gauge groups. After further applying a $\Ztwo$ symmetry, the potential of the model is given by
\begin{eqnarray}\label{eq:potsing}
{ V(\Phi,S ) 
= -m^2 \Phi^{\dagger} \Phi -\mu ^2 S ^2 + \lambda_1
(\Phi^{\dagger} \Phi)^2 + \lambda_2  S^4 + \lambda_3 \Phi^{\dagger}
\Phi S ^2}
\end{eqnarray}
where $\Phi$ and $S$ denote the doublet and singlet in the gauge eigenbasis. Both fields acquire a vacuum expectation value (vev); therefore, mass eigenstates are related by a mixing matrix specified by the angle $\al$. One of the scalar masses as well as the doublet vev are fixed from experimental measurements, such that the model has in total 3 undetermined parameters
\begin{\eqn*}
M_{h/H},\sin\al,\,\tan\be\,\equiv\,\frac{v_\Phi}{v_S},
\end{\eqn*}
where $v$ denote the respective vevs. All couplings to SM particles are inherited from the doublet in the gauge eigenstate, leading to a universal suppression by $\sin\al\,\lb \cos\al \rb$ for $h\,(H)$. We here use the convention that $M_H\,\geq\,M_h$.\\

\begin{center}
\begin{figure}
\begin{center}
\includegraphics[width=0.6\textwidth]{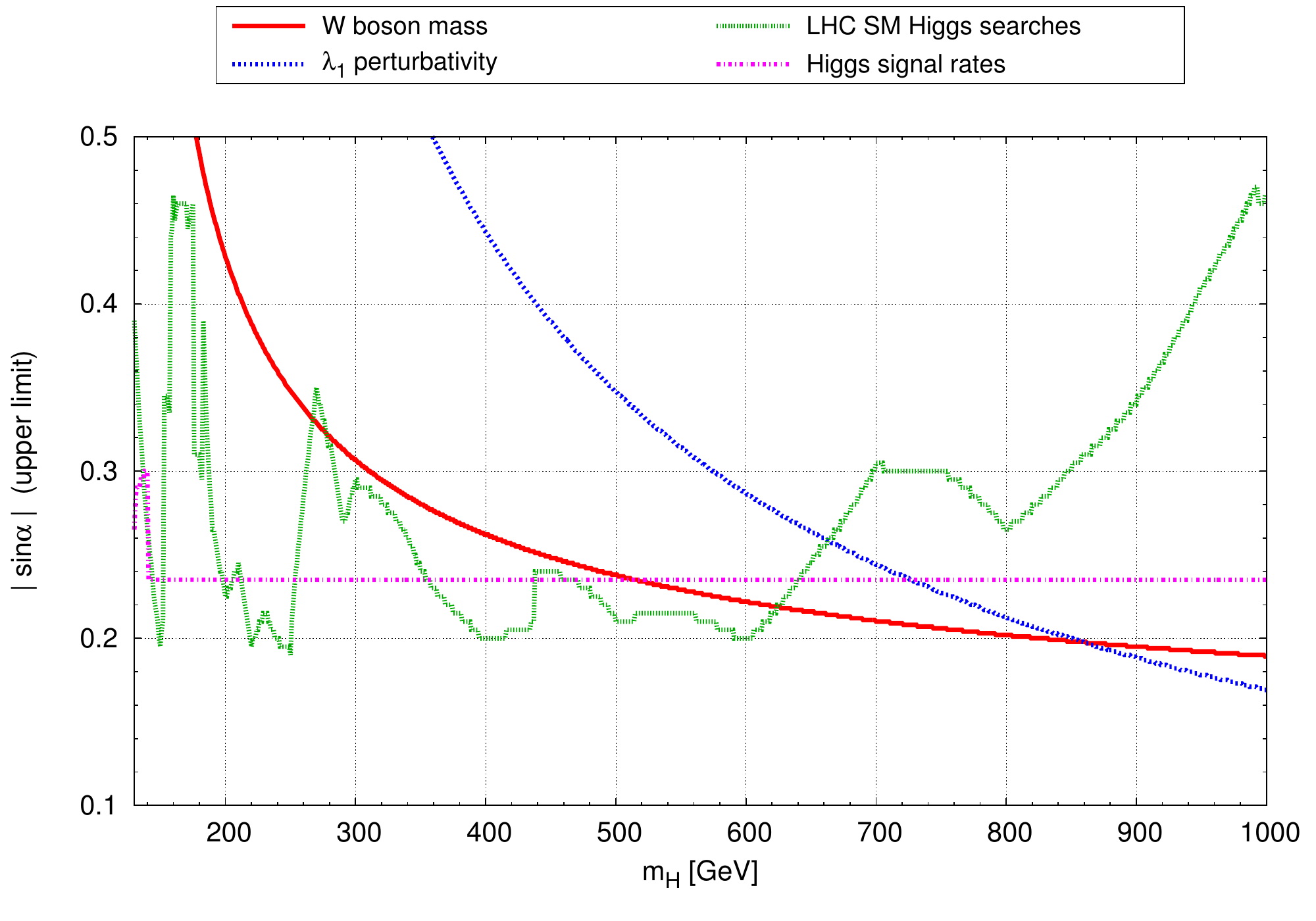}
\caption{\label{fig:singlim} Current constraints on the parameter space of the Higgs singlet extension. Shown are limits from the W-boson mass as a precision observable \cite{Lopez-Val:2014jva} (solid, red), direct searches as implemented in \HiggsBounds (green, dashed), limits from signal strength measurements as implemented in \HiggsSignals (magenta, dashed), as well as limits from perturbativity of $\lam_1$. $\tan\be$ has been fixed to 0.1. This figure corresponds to an update of results presented in \cite{Ilnicka:2018def}.}
\end{center}
\end{figure}
\end{center}

The model is subject to a large number of theoretical and experimental constraints and has been vastly discussed in the literature (see e.g. \cite{Robens:2015gla,Robens:2016xkb,Ilnicka:2018def} and references therein). For briefness, we here concentrate on the case where $M_h\,=\,125\,\GeV$.
In this case, $|\sin\al|=0$ denotes the SM-like decoupling limit. We show a survey of the interplay of all current constraints within this model in figure \ref{fig:singlim}. For large regions of the available parameter space, constraints from the Higgs signal strength measurements as well as direct searches prove to be dominant. For larger masses, however, the use of the W-mass as a precision observable \cite{Lopez-Val:2014jva} as well as perturbativity of the couplings renders the strongest constraints. Note that the strongest experimental limits mainly stem from searches for $pp\,\rightarrow\,H\,\rightarrow\,V V'$ , as well as other measurements \cite{CMS-PAS-HIG-12-045,CMS-PAS-HIG-13-003,Khachatryan:2015cwa,Aaboud:2017rel,Sirunyan:2018qlb,Aaboud:2018bun}.

\begin{figure}
\begin{center}
\includegraphics[width=0.45\textwidth]{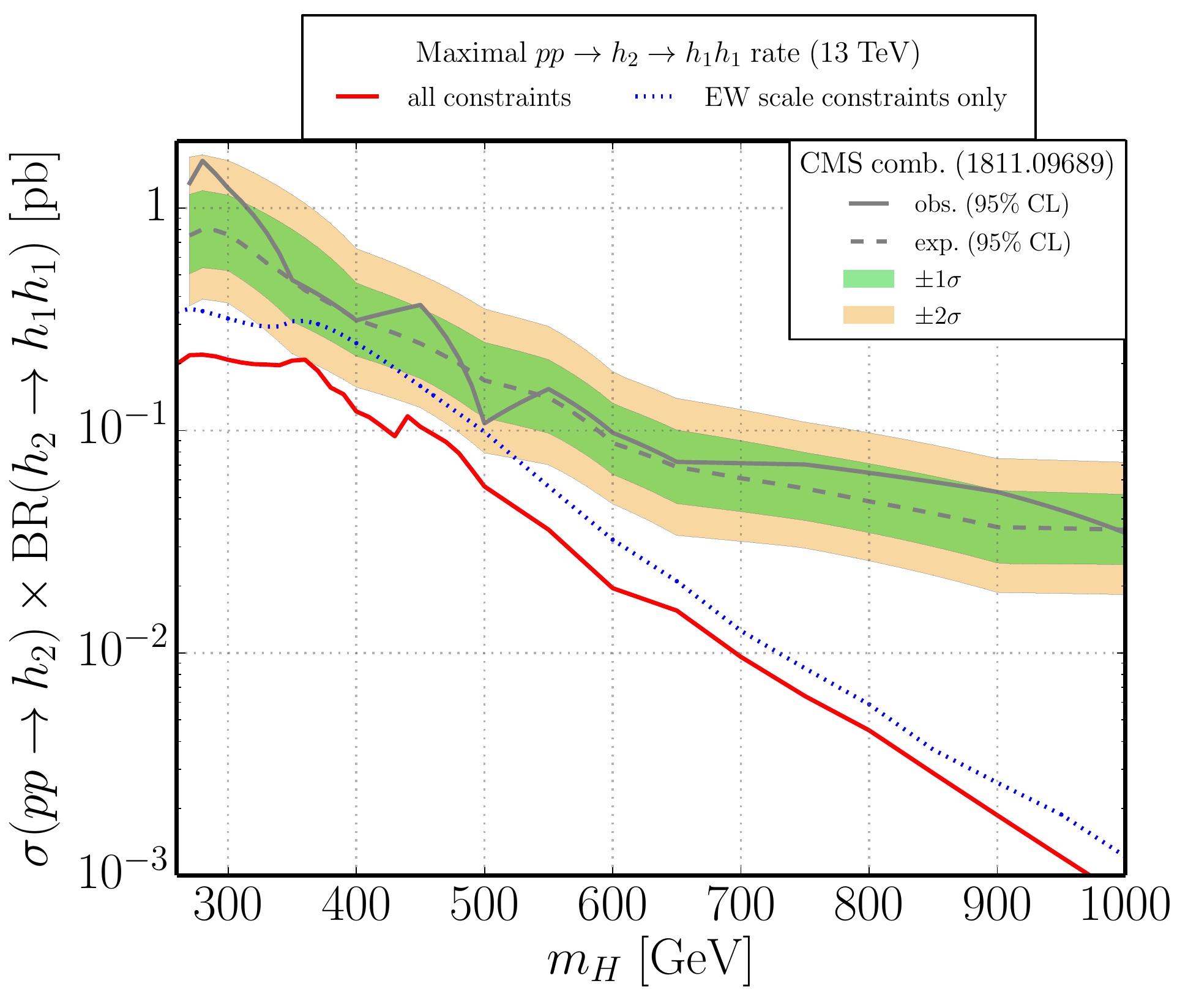}
\includegraphics[width=0.45\textwidth]{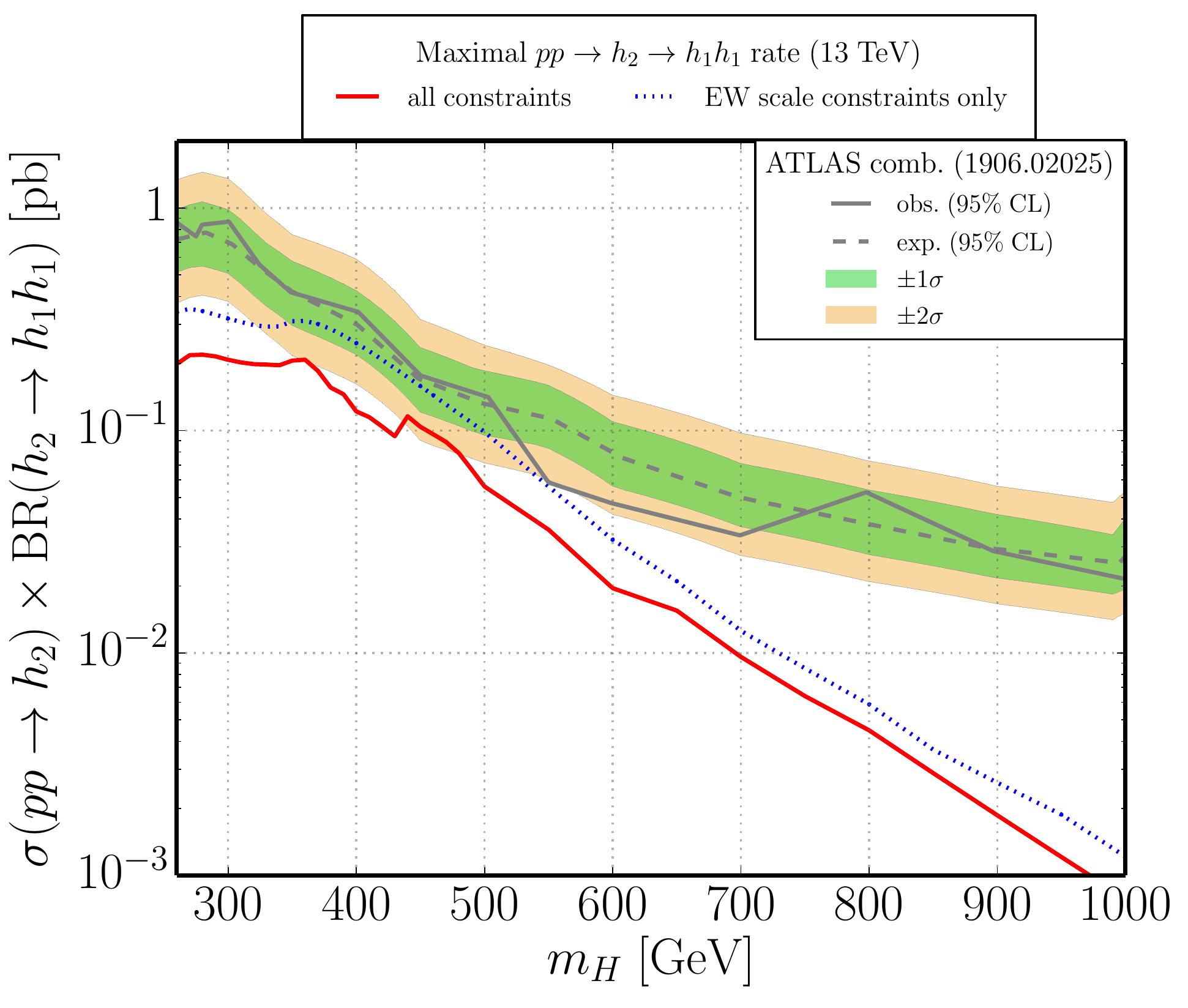}
\caption{\label{fig:xshigh_hh} {Maximal allowed $pp \to h_2 \to h_1h_1$ signal rate at the 13 {\rm TeV}~LHC in the softly-broken $\Ztwo$-symmetric case. Shown are values after applying (red solid) all constraints and (blue dotted) only constraints at the electroweak (EW) scale. For comparison we include the current strongest cross section limit (at $95\%~\mathrm{CL}$), obtained from the combination of various CMS $h_2\to h_1h_1$ searches at $13~\mathrm{TeV}$ with up to $36~\mathrm{fb}^{-1}$ of data~\cite{Sirunyan:2018two} {{\sl (left)} as well as ATLAS Run II searches with the same integrated luminosity \cite{Aad:2019uzh} {\sl (right)}.}} This figure is an update of a result presented in \cite{z2bms}. }
\end{center}
\end{figure}

Another interesting feature is the resonant production of $H$, followed by a subsequent decay into $h\,h$. In fact, this signatures has been vastly investigated by the LHC experiments. We show a comparison of the current experimental search limits \cite{Sirunyan:2018two,Aad:2019uzh} with the maximal rate allowed by all theoretical and experimental constraints in figure \ref{fig:xshigh_hh}. Note that there is an interesting interplay between limits from perturbativity of the couplings and direct searches for $M_H\,\gtrsim\,400\,\GeV$, leading to a decrease of the maximally allowed mixing angle from the value derived from signal strength measurements, $|\sin\al|\,\lesssim\,0.235$, in that mass range.

\section{Inert Doublet Model}
\subsection{Model setup and current constraints}
I now turn to a model that, apart from extending the scalar sector of the SM by an additional particle content, also provides a dark matter candidate. The Inert Doublet Model (IDM) \cite{Deshpande:1977rw,Cao:2007rm,Barbieri:2006dq} is a two-Higgs doublet model that comes with an additional exact $\mathbb{Z}_2$ symmetry. This allows to define different transformation properties of the two doublets, which are labelled $\Phi_S$ and $\Phi_D$, such that

  \begin{equation}
    \Phi_D \rightarrow - \Phi_D, \,\,
    \Phi_S\rightarrow \Phi_S, \,\,
    \text{SM} \rightarrow \text{SM}. \label{eq:dsym}
  \end{equation}
Following the above transformation properties, only the SM-like doublet $\Phi_S$ acquires a vacuum expectation value; therefore, electroweak symmetry breaking proceeds as in the SM. Similarly, the second doublet does not couple to fermions, and therefore all interactions from particle of the second doublet stem from direct scalar-scalar interactions terms in the potential or from the covariant derivative required by gauge symmetry. The model has been vastly explored in the literature (see e.g. \cite{Ilnicka:2015jba,Abe:2018bpo,Ilnicka:2018def,Kalinowski:2018ylg,Dercks:2018wch} and references therein, as well as
\cite{Dolle:2009ft,Swiezewska:2012eh,Gustafsson:2012aj,Arhrib:2012ia,Krawczyk:2013jta,Belanger:2015kga,Ilnicka:2015jba,deFlorian:2016spz,Poulose:2016lvz,Datta:2016nfz,Kanemura:2016sos,Akeroyd:2016ymd, Ilnicka:2018def,Wan:2018eaz,Belyaev:2018ext,Dercks:2018wch,Bhardwaj:2019mts} for studies of LHC phenomenology). 
Signatures at colliders always render electroweak gauge bosons and missing transverse energy. Although the model therefore features collider signatures similar to other models which are currently under investigation at the LHC (see e.g. \cite{Abe:2018bpo} and references therein), the IDM itself has not been explored directly by the LHC experiments.

Due to the above symmetry, $\Phi_S$ plays the role of the SM doublet regarding electroweak symmetry breaking; therefore, the scalar excitation of this doublet, which we denote by $h$, has a mass of $M_h\,=\,125\,\GeV$, and the corresponding vev is $v\,\sim\,246\,\GeV$. The dark scalars from $\Phi_D$ are denoted by $H,\,A\,$ and $H^\pm$; their masses are not determined by direct measurements. Note also that there is no definite CP-quantum number for $A,\,H$ in this scenario. Due to the $\Ztwo$ symmetry, the model contains a stable scalar, which can serve as a dark matter candidate. In the following, we chose $H$ as the DM particle.\\

The potential is then given by
\begin{equation}\begin{array}{c}
V\lb \Phi_S,\,\Phi_D \rb=-\frac{1}{2}\left[m_{11}^2(\Phi_S^\dagger\Phi_S)\!+\! m_{22}^2(\Phi_D^\dagger\Phi_D)\right]+
\frac{\lambda_1}{2}(\Phi_S^\dagger\Phi_S)^2\! 
+\!\frac{\lambda_2}{2}(\Phi_D^\dagger\Phi_D)^2\\[2mm]+\!\lambda_3(\Phi_S^\dagger\Phi_S)(\Phi_D^\dagger\Phi_D)\!
\!+\!\lambda_4(\Phi_S^\dagger\Phi_D)(\Phi_D^\dagger\Phi_S) +\frac{\lambda_5}{2}\left[(\Phi_S^\dagger\Phi_D)^2\!
+\!(\Phi_D^\dagger\Phi_S)^2\right].
\end{array}\label{pot}\end{equation}

After electroweak symmetry breaking, the model has a priori seven free parameters; however, as discussed above, $M_h$ and $v$ are fixed by experimental measurements, leaving in total five free parameters. We chose them as

\begin{\eqn}\label{eq:physbas}
M_H, M_A, M_{H^{\pm}}, \lam_2, \lam_{345}\,\equiv\,\lam_3\,+\lam_4\,+\,\lam_5.
\end{\eqn}

\begin{figure}
\begin{center}
\includegraphics[width=0.49\textwidth]{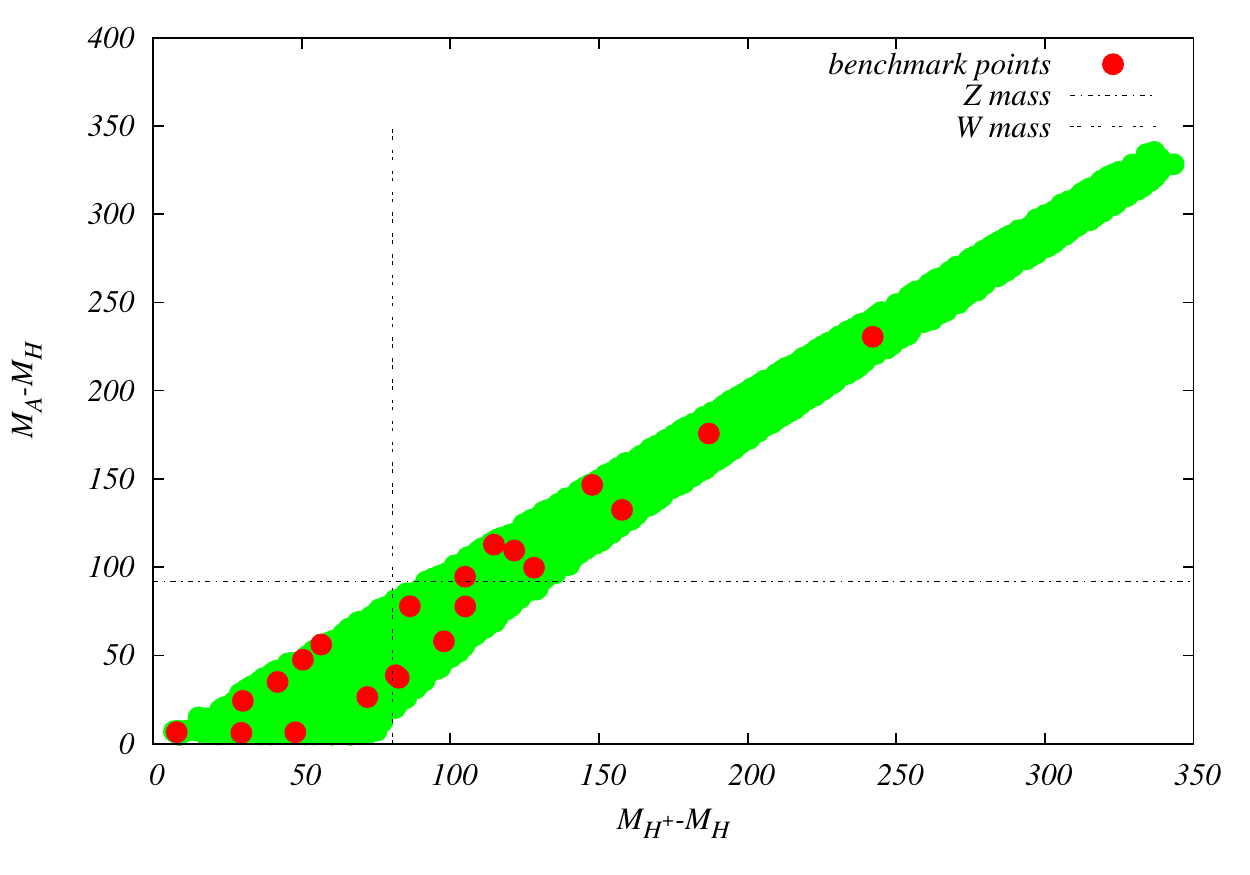}%\\
\includegraphics[width=0.49\textwidth]{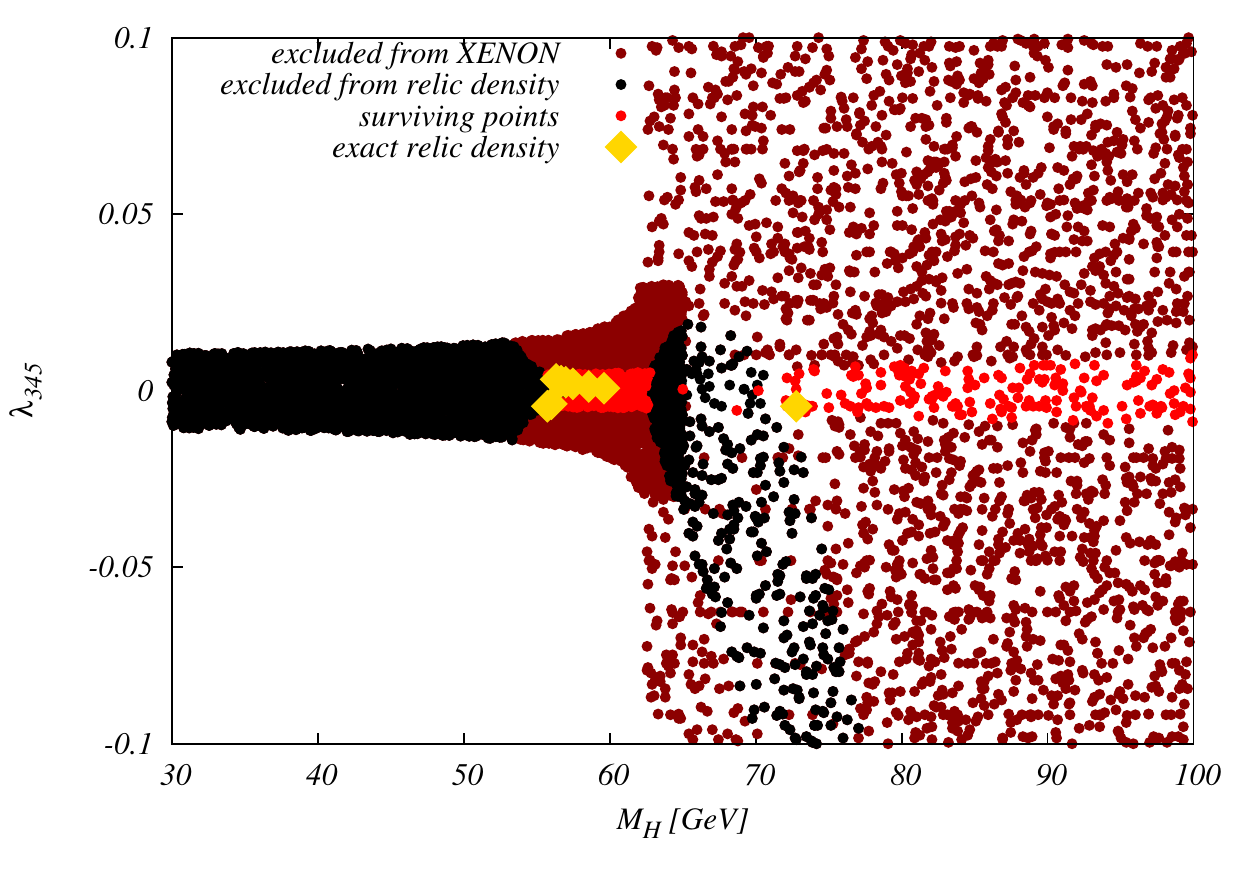}%\\
\end{center}
\caption{\label{fig:idmconst} Allowed parameter space in the IDM. {\sl Left:} Allowed parameter range in mass differences from the generic scan presented in \cite{Kalinowski:2018ylg}, with the figure taken from that reference. Benchmark points chosen in that reference are displayed in red. {\sl Right:} allowed parameter region for the case that $M_H\,\leq\,100\,\GeV$, with allowed parameter points displayed in red. The shape for $M_H\,\leq\,M_h/2$ is determined by signal strength measurements. Dark matter observables have been obtained using \texttt{micrOmegas} \cite{Barducci:2016pcb}. Figure taken from \cite{Ilnicka:2018def}.}
\end{figure}

The model is subject to a large number of theoretical and experimental constraints (see \cite{Ilnicka:2015jba,Ilnicka:2018def,Kalinowski:2018ylg} for a survey and more recent updates). In general, constraints differ for the region where $M_H\,>\,M_h/2$ and $M_H\,<\,M_h/2$; in the latter scenario, extremely strong limits stem from a combination of signal strength measurements of the 125 \GeV~ Higgs and dark matter relic density limits. We exemplify this in figure \ref{fig:idmconst}, where we show the allowed parameter regions for both cases. For $M_H\,\geq\,M_h/2$, relatively strong constraints stem from electroweak precision measurements, parametrized via the oblique parameters $S,T,U$ \cite{Altarelli:1990zd,Peskin:1990zt,Peskin:1991sw,Maksymyk:1993zm}\footnote{The values of the oblique parameters as well as other physical observable predictions have been obtained using  the Two Higgs Doublet Model Calculator (\texttt{2HDMC}) tool \cite{Eriksson:2009ws}.}, and masses of the dark scalars are highly correlated. On the other hand, for $M_H\,\leq\,M_h/2$, an interplay of signal strength measurements and dark matter constraints pose the strongest bounds. 

The model features interesting signatures at colliders. At hadron colliders, 
\begin{\eqn*}
p\,p\,\rightarrow\,H\,A,\;H\,H^\pm,\;A\,H^\pm,\;H^+\,H^-
\end{\eqn*}

while for $e^+ e^-$ colliders we have 

\begin{\eqn*}
e^+ e^-\,\rightarrow\,H\,A,\;H^+\,H^-.  
\end{\eqn*}
The production at both machines is mainly mediated via electroweak (Drell-Yan) processes, where all couplings are determined by SM quantities in the electroweak sector, such that cross sections mainly depend on the available phase space. Production cross sections for the above processes can reach up to $1\,\pb$ at the 13 \TeV~ LHC \cite{trtalk} and $\mathcal{O}\lb 100\,\fb \rb$ at $e^+e^-$ colliders \cite{Kalinowski:2018ylg}. However, these values are highly dependent on the considered parameter point. The unstable dark scalars dominantly decay via
\begin{\eqn*}
A\,\rightarrow\,Z^{(*)} H,\,H^\pm\,\rightarrow\,W^{\pm (*)}\,H
\end{\eqn*}
where the on- or offshellness of the electroweak gauge bosons depends on the kinematic features of the considered parameter point.

\begin{center}
\begin{figure}
\begin{center}
\includegraphics[width=0.49\textwidth]{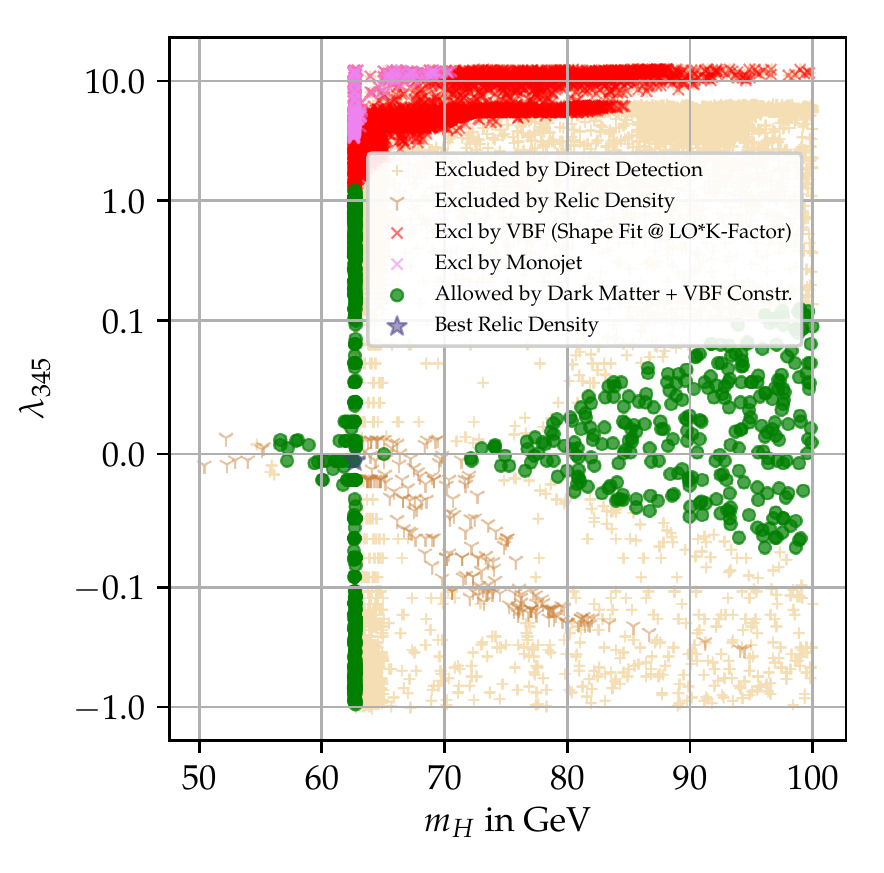}%\\
 \includegraphics[width=0.49\textwidth]{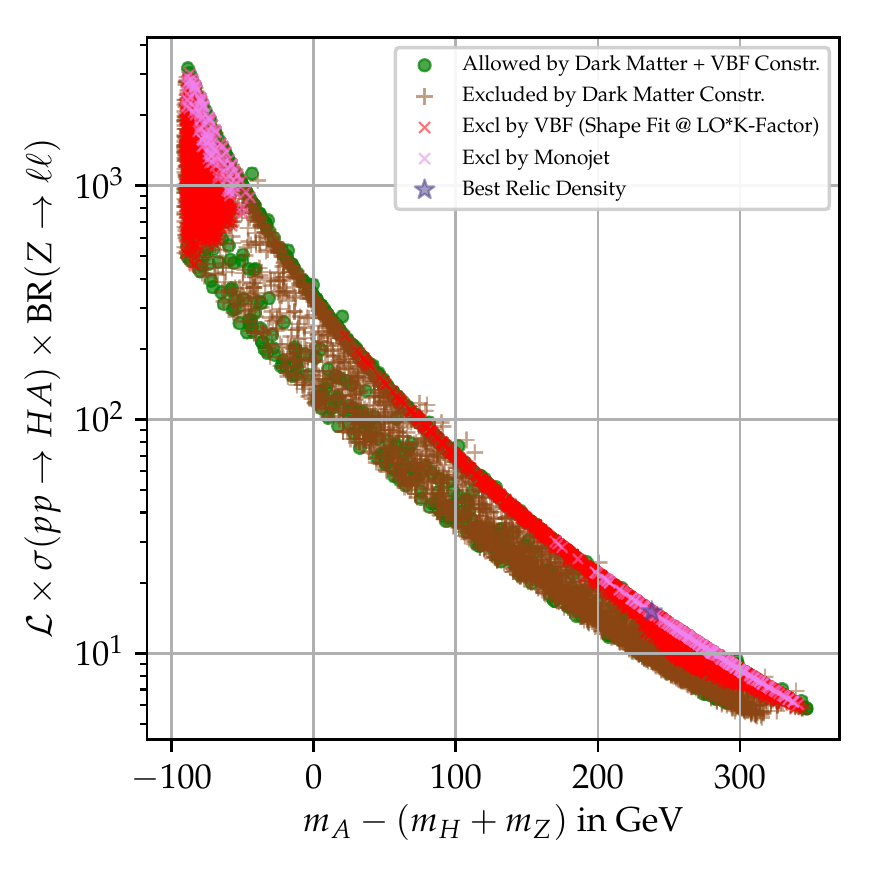}%\\
\caption{\label{fig:idm_recast} Results of the recast of the Inert Doublet Model, presented in \cite{Dercks:2018wch}. {\sl Left:} Constraints on the parameter space in the $(M_H,\,\lam_{345})$ plane taking all theoretical and experimental constraints into account. Recasts of VBF \cite{Sirunyan:2018owy} and monojet \cite{ATLAS-CONF-2017-060,Aaboud:2017phn} searches are also shown. For $M_h\,\gtrsim\,M_h/2$, the VBF recast provides the only constraint in certain regions of parameter space. {\sl Right:} Production cross section for the dilepton and missing transverse energy final state in the IDM, vs. estimate of the available missing transverse energy from kinematic considerations. Cuts on $\slashed{E}_\perp\,\gtrsim\,100\,\GeV$ reduce the production cross sections by roughly an order of magnitude. }
\end{center}
\end{figure}

\end{center}

While the model is not yet directly explored by the LHC experiments, it is interesting to see whether recasts of already existing searches can already cast some light on the possible reach of more dedicated studies. In \cite{Belanger:2015kga}, LHC Run I results in the dilepton channel were used in order to investigate the parameter space of the model; however, most regions that seem to be sensitive to these signatures are already highly constrained by astrophysical measurements. In \cite{Dercks:2018wch}, instead, Run II data for VBF production with of a SM like Higgs with a subsequent invisible decay \cite{Sirunyan:2018owy} was used in order to limit the models parameter space. While regions where $M_H\,<\,M_2/h$ are strongly constrained by measurements of the invisible branching ratio of $h$, with the current best limit being $\text{BR}(h\rightarrow\,\text{invisible})\,\leq\,19\%$\cite{Sirunyan:2018owy}\footnote{Note that the updated result on the combined limit has been made available in May 2019.}, the region where $H$ is slightly offshell can be further constrained by a reinterpretation of the corresponding experimental analysis; in general, this recast limits an upper triangle in the $\lb M_H,\,\lam_{345}\rb$ plane, see figure \ref{fig:idm_recast}. This emphasizes the importance of comparing fully simulated samples, including all offshell and interference effects, with current limits from collider searches.

In addition, in \cite{Dercks:2018wch} it was demonstrated that current 13 \TeV~ multi-lepton searches are not sensitive to the IDM within the allowed parameter regimes. This can be traced back to relatively high cuts on missing transverse energy $\slashed{E}_\perp$; figure \ref{fig:idm_recast} shows that there is an almost log-like decay of the production rates as a function of the available $\slashed{E}_\perp$ per parameter point. However, an adjustment of the corresponding cuts could in principle open up the according parameter space, enhancing the corresponding signal rates by nearly an order of magnitude. Especially for leptonic final states, this should not pose an impediment on reasonable trigger requirements.

The above example shows that there are viable theoretical models that are well-studied from the theoretical viewpoint and render well-known experimental signatures, that are yet under-investigated by the LHC experiments. I strongly encourage the experimental collaborations to widen their searches to take such models, including modified cut strategies, into account.

\subsection{Future collider prospects}

\begin{center}
\begin{figure}
\begin{center}
\includegraphics[width=0.45\textwidth]{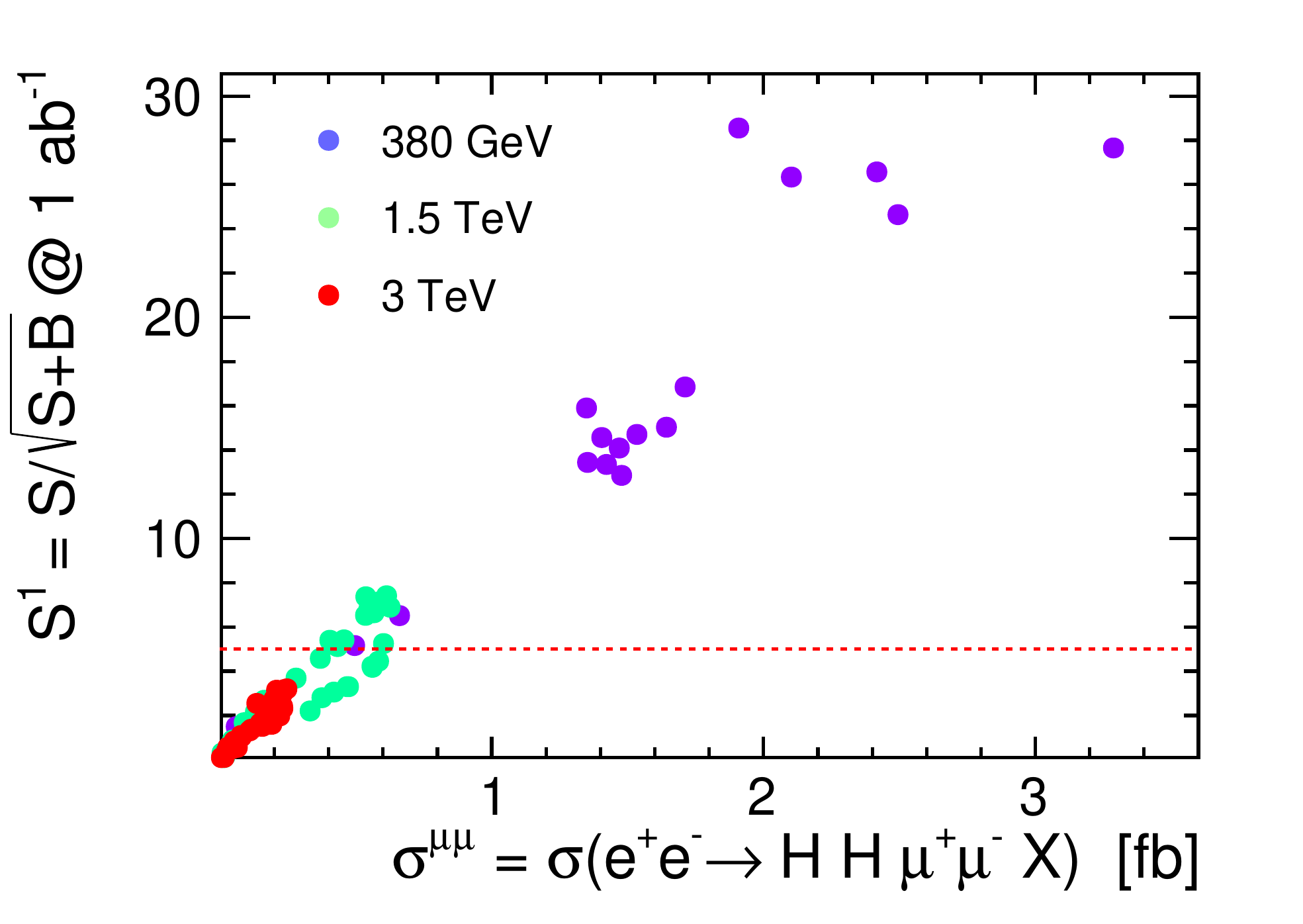}
\includegraphics[width=0.45\textwidth]{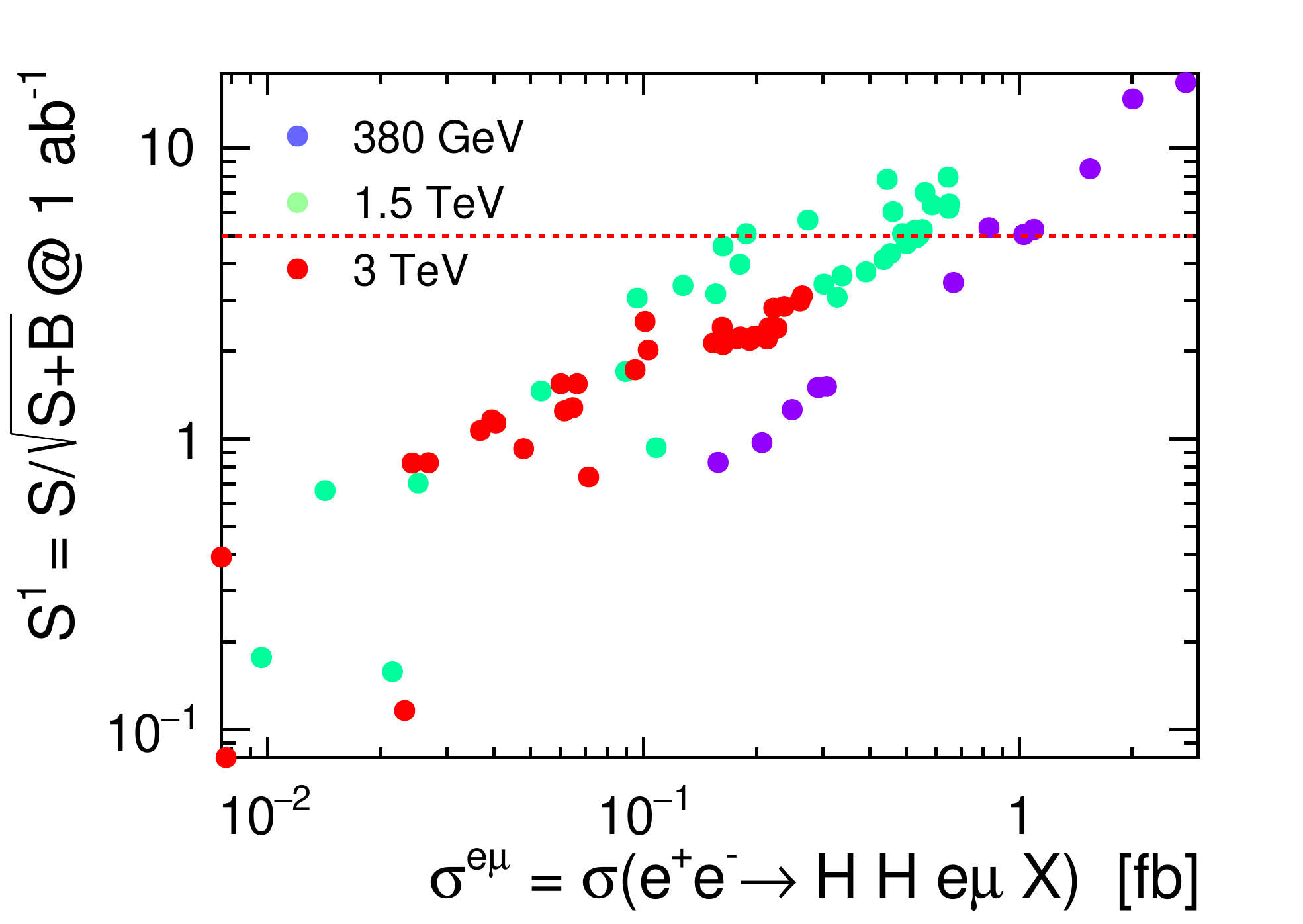}
\includegraphics[width=0.45\textwidth]{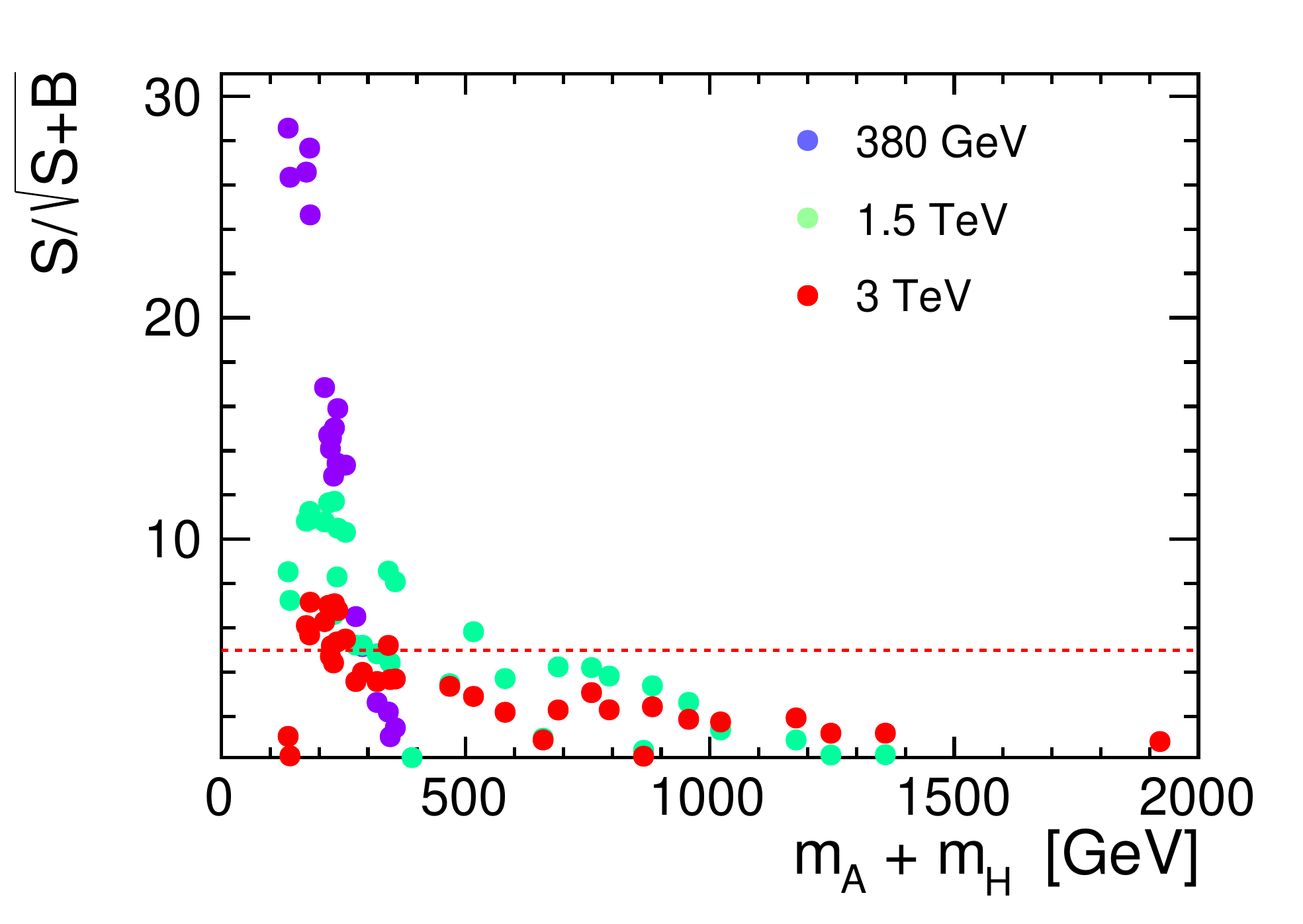}
\includegraphics[width=0.45\textwidth]{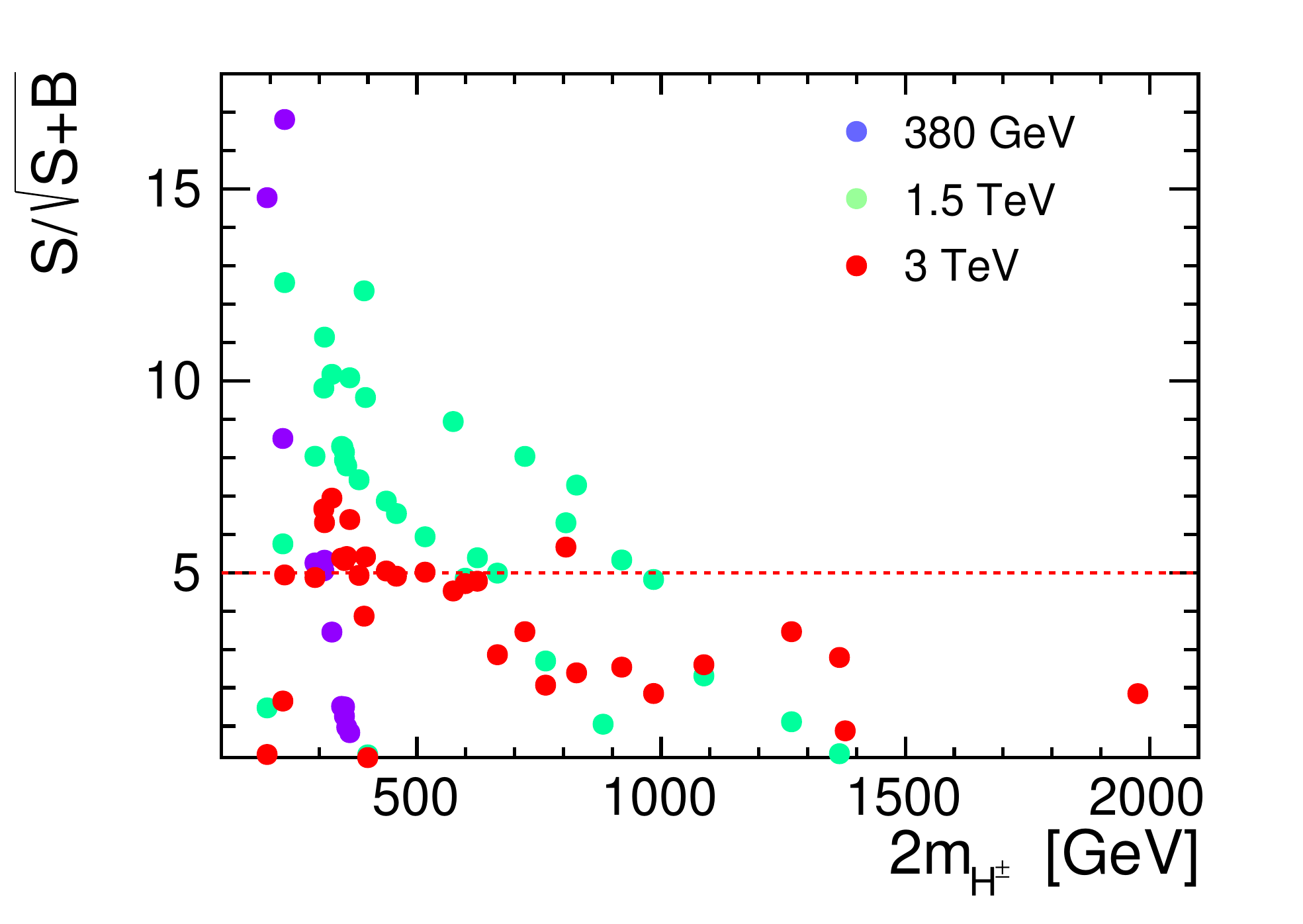}
\end{center}
\caption{\label{fig:idm_clic} Significances obtained for $e^+\,e^-\,\rightarrow\,\ell\ell'+\slashed{E}_\perp$ final state within the IDM for various center-of-mass energies, resulting from dedicated studies presented in \cite{Kalinowski:2018kdn}. {\sl Left column:} significances in the $\ell\ell'\,\equiv\,\mu^+\mu^-$ channel, with major production mode via $HA$ production. {\sl Top:} Significance at $\int\mathcal{L}\,=\,1\,\ab^{-1}$, for various center-of-mass energies. In general, discovery is possible if $\sigma(e^+e^-\,\rightarrow\,\mu^+\mu^-+\slashed{E}_\perp)\,\gtrsim\,1\fb$ prior to cuts. {\sl Bottom:} Significances as a function of the mass scale $M_A+M_H$, with respective integrated target luminosities of $\int\mathcal{L}\,=\,\lb 1, 2.5, 5\rb\,\ab^{-1}$ for $\sqrt{s}\,=\,380\,,\GeV, 1.5\,, 3\,\TeV$.  {\sl Right column, top and bottom:} significances in the $\ell\ell'\,\equiv\,e^\pm\mu^\mp$ channel, following mainly $H^+\,H^-$ production. The mass scale in the bottom plot is now given by $2\,M_{H^\pm}$. }
\end{figure}
\end{center}

Benchmarks for the investigation of the IDM at $e^+\,e^-$ colliders, taking all current constraints into account, have been proposed in \cite{Kalinowski:2018ylg}. Results, based on dedicated simulation of all signal and background processes, with beamstrahlung and detector acceptance taken into account, have been presented in \cite{Kalinowski:2018kdn,deBlas:2018mhx} for CLIC running at $380\,\GeV,\,1.5\,\TeV$ and $3\,\TeV$ (see \cite{Zarnecki:2019poj} for an extension of this work to center-of-mass energies of $250$ and $500\,\GeV$). The analysis was based on a combination of pre-cuts with the application of boosted decision trees. In general, significances up to $\sim\,30$ can be obtained using these methods, where best results were obtained for mass scales\footnote{Mass scale denotes the sum of the produced dark scalar pair masses.} below $500\,\GeV$ at $\sqrt{s}\,=\,380\,\GeV$. We investigated $\mu^+\,\mu^-\,+\,\slashed{E}_\perp$ as well as $e^\pm\mu^\mp+\slashed{E}_\perp$ final states and found that in general an approximate limit of $\sigma\lb e^+\,e^-\,\rightarrow\,\ell\,\ell' +\slashed{E}_\perp\rb\,\sim\,1\fb$  can be taken as a lower limit above which discovery seems feasible for integrated luminosities $\int\mathcal{L}\,=\,1\ab^{-1}$. The corresponding results are shown in figure \ref{fig:idm_clic}\footnote{Note that investigations in the semi-leptonic channel can improve the mass reach by a factor 2; cf. e.g. \cite{dorotatalk} for a presentation of preliminary results.}. Model implementation and event generation were done making use of a SARAH \cite{Staub:2015kfa}, {\tt SPheno 4.0.3} \cite{Porod:2003um,Porod:2011nf}, and {\tt WHizard 2.2.8} \cite{Moretti:2001zz,Kilian:2007gr}. Routines for boosted decision trees were obtained from the TMVA toolkit \cite{Hocker:2007ht}.

In general, therefore, BSM scenarios where production modes are mainly goverend by electroweak processes also provide an important set of models that can be tested at future $e^+ e^-$ colliders down to relatively low production cross sections.

\section{Two real scalar extension at hadron colliders}
Aside from already explored signatures that might be reinterpreted in or modified for different new physics models, another important topic are signatures which have so far not been explored by the experimental collaborations using the currently available dataset. A prominent example for this are scalar-to-scalar decays. While searches for new physics states that decay into the SM scalar, as well as the decay of the SM scalar into two lighter particles have already been investigated (see e.g. \cite{Aaboud:2018iil,Aaboud:2018esj,Sirunyan:2018mot,Sirunyan:2019gou} for recent searches), single scalar production following by asymmetric scalar decays or symmetric decays, where all scalar masses differ from the SM-like scalar mass at 125 \GeV, have not yet been fully explored. Simple counting reveals that for such scenarios at least three physical scalar states need to be present in the model, out of which one takes the role of the state already discovered by the LHC experiments.

The simplest example for such a new physics model is described by the extension of the SM scalar sector by two real (or one complex) scalar(s). The most general model contains in total 21 parameters in the scalar sector \cite{Barger:2008jx}; however, additional symmetries can be imposed that lead to a dramatic decrease of the number of free parameters (see e.g. \cite{Coimbra:2013qq,Costa:2015llh}).

We here discuss a model that obeys the $\Ztwo\,\otimes\,\Ztwo'$ symmetry
$        \Ztwo^S: \, S\to -S\eqcomma
        \Ztwo^X: \, X\to -X,$ while all other fields transform evenly under the respective $\Ztwo$ symmetry.
This model has recently been discussed in great detail in \cite{Robens:2019kga}. The potential in the scalar sector is then given by
\begin{equation}
    \begin{aligned}
        V\lb \Phi,\,S,\,X\rb & = \mu_{\Phi}^2 \Phi^\dagger \Phi + \lambda_{\Phi} {(\Phi^\dagger\Phi)}^2
        + \mu_{S}^2 S^2 + \lambda_S S^4
        + \mu_{X}^2 X^2 + \lambda_X X^2                                              \\
          & \quad+ \lambda_{\Phi S} \Phi^\dagger \Phi S^2
        + \lambda_{\Phi X} \Phi^\dagger \Phi X^2
        + \lambda_{SX} S^2 X^2\eqdot
    \end{aligned}\label{eq:TRSMpot}
\end{equation}
where $\Phi$ denotes the doublet also present in the SM potential and $X,\,S$ are two additional real scalars that are singlets under the SM gauge groups. All three scalars acquire a vev and mix. This leads to three physical states with all possible scalar-scalar interactions.

The model, including all theoretical and experimental constraints, has been described in \cite{Robens:2019kga}, which we refer the reader to for further reference. Main constraints stem from the Higgs signal strength measurements by the LHC experiments. We also found that perturbative unitarity as well as boundedness from below pose important constraints, in addition to current collider searches. Results have been obtained using a private version of the \texttt{ScannerS}~\cite{Coimbra:2013qq,Ferreira:2014dya,Costa:2015llh,Muhlleitner:2016mzt} framework. 

In the following, we will use the convention that
\begin{\eqn}\label{eq:hier}
M_1\,\leq\,M_2\,\leq\,M_3
\end{\eqn}
and denote the corresponding physical mass eigenstates by $h_i$.
Gauge and mass eigenstates are related via a mixing matrix. Interactions with SM particles are then inherited from the scalar excitation of the doublet via rescaling factors $\kappa_i$, such that $g_i^{h_i A B}\,=\,\kappa_i\,g_i^{h_i A B,\text{SM}}$ for any $h_i A B$ coupling, where $A,\,B$ denote SM particles. Orthogonality of the mixing matrix implies $\sum_i \kappa_i^2\,=\,1$. Furthermore, signal strength measurements require $|\kappa_{125}|\gtrsim\,0.96$ \cite{Robens:2019kga} for the SM-like scalar $h_{125}$, which can be $h_1,\,h_2$ or $h_3$ depending on the specific parameter choice.

The model features interesting scalar-to-scalar decays
\begin{align}
    pp\rightarrow h_a~(+ X)\rightarrow h_b h_b~(+ X),  \label{eq:process_sym} \\
    pp\rightarrow h_3~(+ X)\rightarrow h_1 h_2~(+ X), \label{eq:process_asym}
\end{align}
where ${a,b}\,\in\,\left\{1,2,3 \right\}$ under the assumption that the mass hierarchy (\ref{eq:hier}) is fulfilled and on-shell decays are kinematically allowed. In \cite{Robens:2019kga}, six benchmark planes (BPs) were suggested for the decay chains above: three involving asymmetric decays (\ref{eq:process_asym}) and three symmetric decays (\ref{eq:process_sym}), where in the latter case only scenarios were considered where $M_{a,b}\,\neq\,125\,\GeV$. Depending on the specific benchmark point, production rates can reach $\mathcal{O}\lb 60\,\pb \rb$ for the above production modes. The benchmark scenario which considers $h_3\,\rightarrow\,h_2\,h_2$, where $M_2\,>\,250\,\GeV$, can furthermore lead to interesting $h_{125}h_{125}h_{125}$ and $h_{125}h_{125}h_{125}h_{125}$ final states. Largest rates for these processes, including further decays to SM final states, are in the $\mathcal{O}\lb \fb \rb$ range.

\begin{center}
\begin{figure}
\begin{center}
\includegraphics[width=0.45\textwidth]{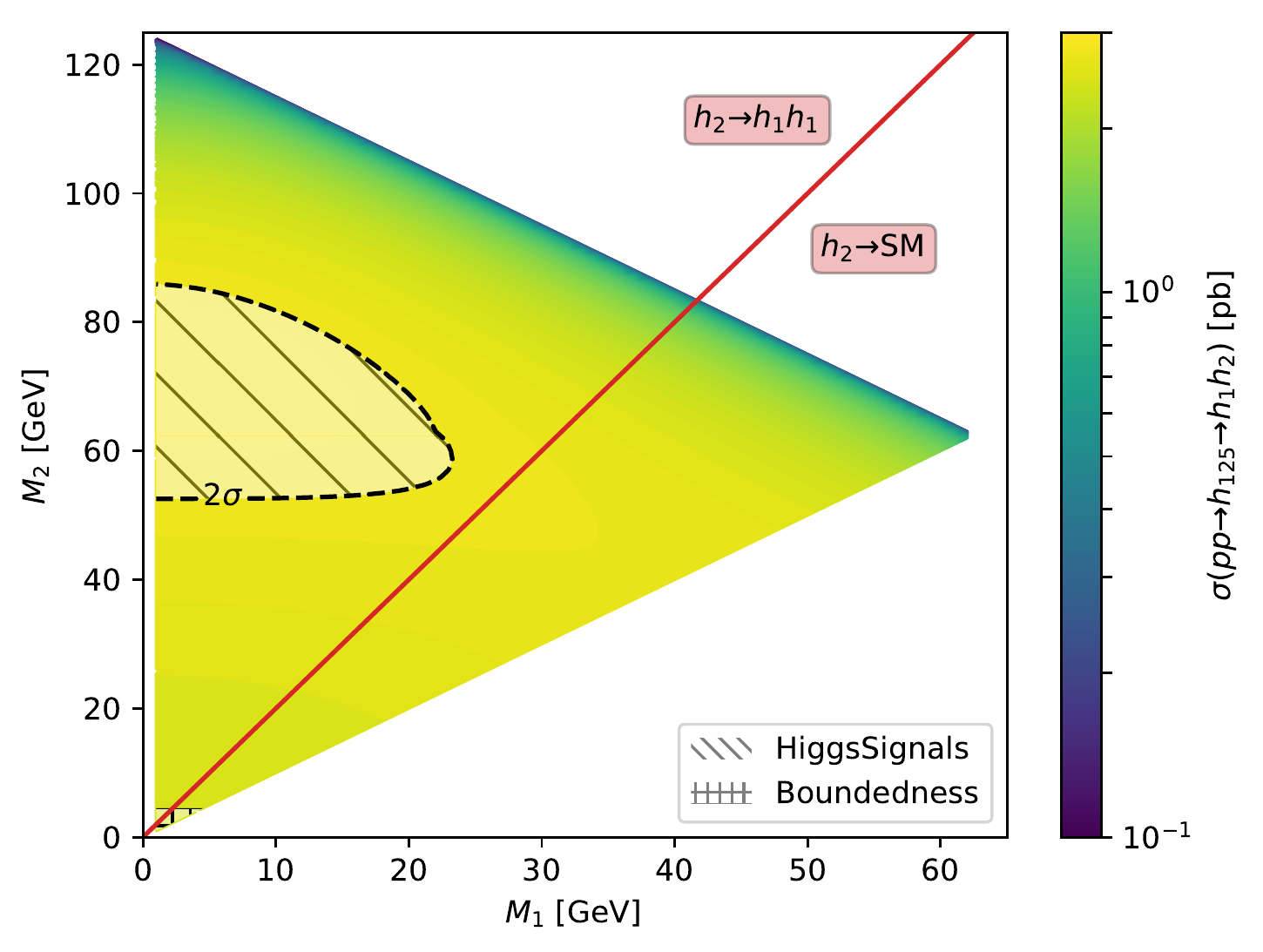}
\includegraphics[width=0.45\textwidth]{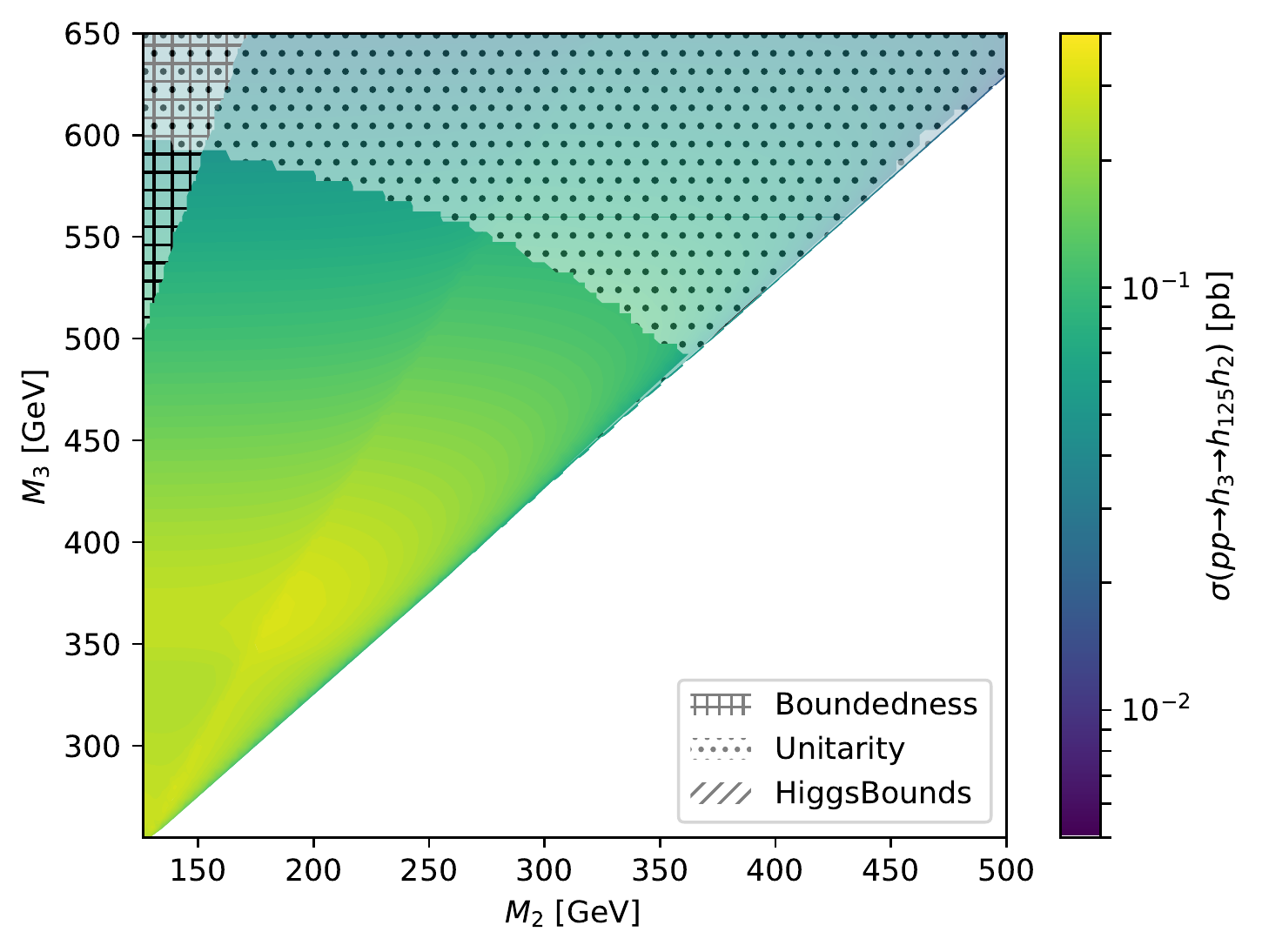}
\includegraphics[width=0.45\textwidth]{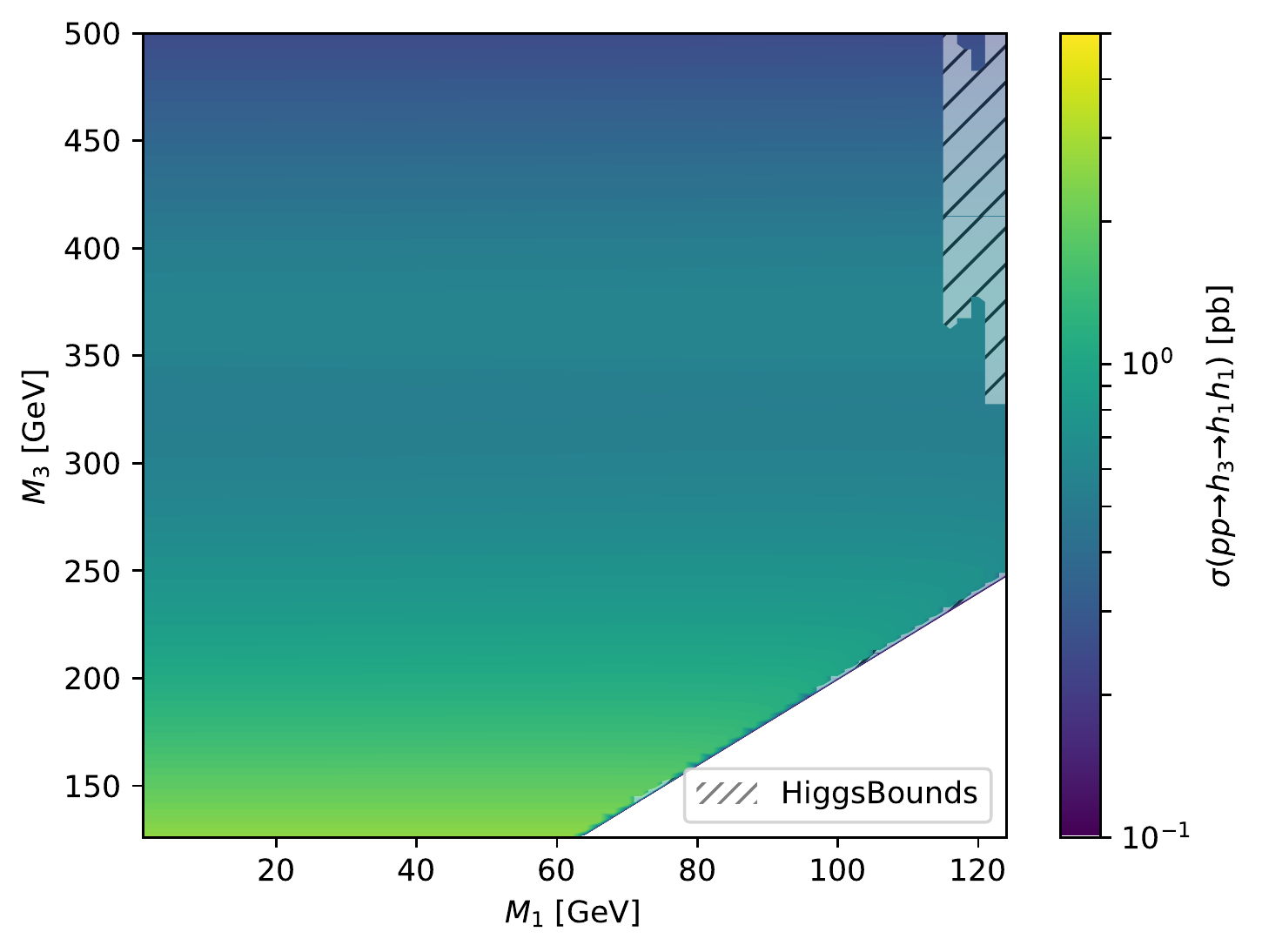}
\includegraphics[width=0.45\textwidth]{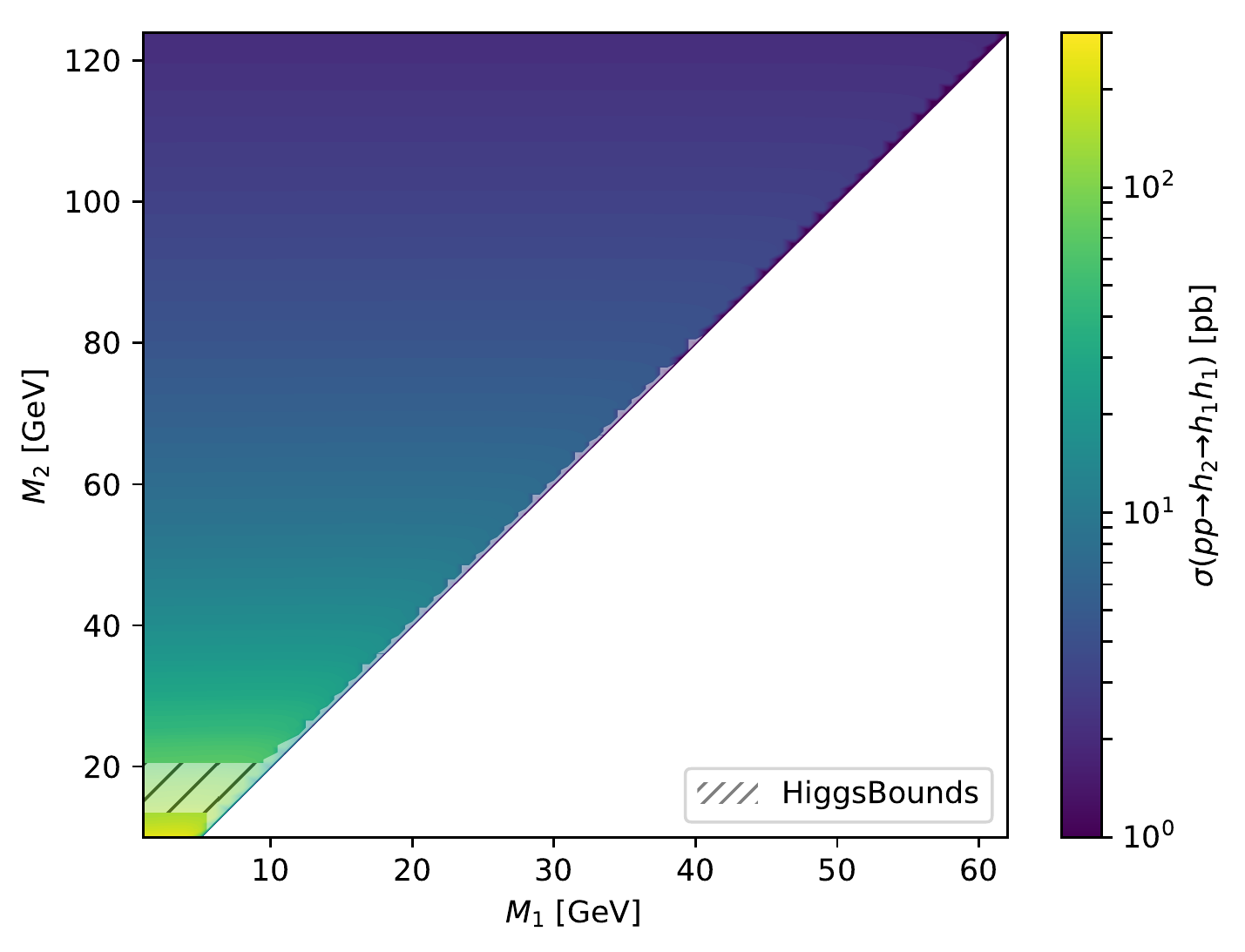}
\end{center}
\caption{\label{fig:2rbps} Signal rate predictions for various benchmark planes suggested in \cite{Robens:2019kga}, at the 13 \TeV~ LHC. Regions forbidden by specific theoretical or experimental constraints are indicated accordingly. {\sl Top left:} BP1, where $M_3\,=\,125\,\GeV$. Production rates reach $\sim\,3\,\pb$, with dominant branching ratios into $b\bar{b}b\bar{b}$ and $b\bar{b}b\bar{b}b\bar{b}$ final states. {\sl Top right:} BP3, where $M_1\,=\,125\,\GeV$. Production rates reach $\sim\,0.3\,\pb$. Dominant branching ratios are to $b\bar{b}W^+W^-$ and $b\bar{b}W^+W^-W^+W^-$, depending on the region in parameter space. {\sl Bottom left:} BP5, where $M_2\,=\,125\,\GeV$. Production rates reach up to $3\,\pb$. Dominant decays are into $b\bar{b}b\bar{b}$ final states. {\sl Bottom right:} BP4, where $M_3\,=\,125\,\GeV$. Signal rates can reach up to $60\,\pb$, with dominant decays again leading to $b\bar{b}b\bar{b}$ final states.}
\end{figure}
\end{center}

Figure \ref{fig:2rbps} shows predicted production rates for several of these benchmark planes: BPs 1/3 for the asymmetric decay, where $h_{125}\,\equiv\,h_3/ h_1$, as well as two symmetric scenarios BPs 4/ 5 where a non-SM like scalar decays into light scalars with $M_1\,\leq\,125\,\GeV$. Depending on the benchmark point, dominant final states typically contain multiple b-jets $b\bar{b}b\bar{b}\lb b\bar{b}\rb$. It should also be noted that a small region of parameter space for the symmetric decay $h_3\,\rightarrow\,h_2\,h_2$ has recently been investigated by the ATLAS experiment in the $W^+ W^- W^+ W^-$ channel \cite{Aaboud:2018ksn}.

The model introduced above renders the simplest extension of SM scalar sector that allows for the new processes (\ref{eq:process_sym}), (\ref{eq:process_asym}). Our recent study has shown that these can lead to sizeable rates at the 13 \TeV~ LHC, while still complying with all current theoretical and experimental bounds. I strongly encourage the experimental collaborations to (further) investigate such scenarios using both current as well as future LHC data.

\section{Conclusion}
After the discovery of a scalar which so far complies with predictions for a SM Higgs particle within the theoretical and experimental uncertainties, an important quest for the theoretical and experimental community is to establish whether the discovered particle corresponds to the scalar predicted by the SM, or whether it is part of a more extended scalar sector. I here commented on several scenarios introducing new particle content. The first extends the SM scalar sector by a real scalar which is a singlet under the SM gauge group. This model features a minimal set of additional parameters and therefore provides an ideal testbed to understand the interplay between theoretical and experimental constraints. For this model, direct searches as well as measurements of the Higgs signal strength start to dominate the constraints on the parameter space up to scalar masses $M_H\,\lesssim\,1\,\TeV$, while electroweak precision measurements as well as theoretical bounds also provide important constraints in some of the parameter space. In addition, I discussed the Inert Doublet Model, a two Higgs model that provides a dark matter candidate. This model has so far not directly been explored by the LHC experiments. Although production rates could in principle lead to a sizeable number of new physics particles already within the current dataset, current searches for similar final states would have to be modified in order to increase sensitivity for this model. In contrast, dedicated studies of this model at future $e^+e^-$ facilities indicate that several signatures could be explored at such machines, with mass scales $\,\lesssim\,500\,\GeV$ being accessible using the leptonic channel. For high energy CLIC stages, this reach is expected to be extended up to about 1 TeV when the semi-leptonic channel is considered. Similar dedicated experimental studies should be performed within a hadron collider framework. Finally, I have commented on a model that renders novel scalar-to-scalar decays, with signatures so far not extensively explored by the experimental community. These symmetric or asymmetric decays discussed here can render relatively high cross sections, up to $\mathcal{O}\lb 60\,\pb \rb$ depending on the chosen parameter point and benchmark scenario. These certainly should be explored in more detail within the near future.\\
In summary, while the experimental collaborations have already constrained a larger number of models with extended scalar sectors in many dedicated experimental studies, many open questions still remain. It is imminent that the currently available Run-II dataset should be explored for the modified or new searches discussed here, and that insight from such studies should be used to further design search strategies at future colliders. In addition, it is important to investigate the discovery potential of future $e^+\,e^-$ colliders for these models.

\section{Acknowledgements}
This research was supported in parts by the National Science Centre,
Poland, the HARMONIA project under contract UMO-2015/18/M/ST2/00518
(2016-2019) and OPUS project under contract UMO-2017/25/B/ST2/00496
(2018-2021) and by the European Union through the Programme Horizon 2020 via the COST action CA15108 - Connecting insights in fundamental physics (FUNDAMENTALCONNECTIONS).

\providecommand{\href}[2]{#2}\begingroup\raggedright\endgroup


\begin{thebibliography}{10}

\bibitem{atlpub}
ATLAS public results, https://twiki.cern.ch/twiki/bin/view/AtlasPublic.

\bibitem{cmspub}
CMS public results,
  https://twiki.cern.ch/twiki/bin/view/CMSPublic/PhysicsResults.

\bibitem{Aad:2012tfa}
{\scshape ATLAS} collaboration, \emph{{Observation of a new particle in the
  search for the Standard Model Higgs boson with the ATLAS detector at the
  LHC}}, \href{https://doi.org/10.1016/j.physletb.2012.08.020}{\emph{Phys.
  Lett.} {\bfseries B716} (2012) 1}
  [\href{https://arxiv.org/abs/1207.7214}{{\ttfamily 1207.7214}}].

\bibitem{Chatrchyan:2012xdj}
{\scshape CMS} collaboration, \emph{{Observation of a New Boson at a Mass of
  125 GeV with the CMS Experiment at the LHC}},
  \href{https://doi.org/10.1016/j.physletb.2012.08.021}{\emph{Phys. Lett.}
  {\bfseries B716} (2012) 30}
  [\href{https://arxiv.org/abs/1207.7235}{{\ttfamily 1207.7235}}].

\bibitem{Schabinger:2005ei}
R.~M. Schabinger and J.~D. Wells, \emph{{A Minimal spontaneously broken hidden
  sector and its impact on Higgs boson physics at the large hadron collider}},
  \href{https://doi.org/10.1103/PhysRevD.72.093007}{\emph{Phys. Rev.}
  {\bfseries D72} (2005) 093007}
  [\href{https://arxiv.org/abs/hep-ph/0509209}{{\ttfamily hep-ph/0509209}}].

\bibitem{Patt:2006fw}
B.~Patt and F.~Wilczek, \emph{{Higgs-field portal into hidden sectors}},
  \href{https://arxiv.org/abs/hep-ph/0605188}{{\ttfamily hep-ph/0605188}}.

\bibitem{Deshpande:1977rw}
N.~G. Deshpande and E.~Ma, \emph{{Pattern of Symmetry Breaking with Two Higgs
  Doublets}}, \href{https://doi.org/10.1103/PhysRevD.18.2574}{\emph{Phys. Rev.}
  {\bfseries D18} (1978) 2574}.

\bibitem{Barbieri:2006dq}
R.~Barbieri, L.~J. Hall and V.~S. Rychkov, \emph{{Improved naturalness with a
  heavy Higgs: An Alternative road to LHC physics}},
  \href{https://doi.org/10.1103/PhysRevD.74.015007}{\emph{Phys.Rev.} {\bfseries
  D74} (2006) 015007} [\href{https://arxiv.org/abs/hep-ph/0603188}{{\ttfamily
  hep-ph/0603188}}].

\bibitem{Cao:2007rm}
Q.-H. Cao, E.~Ma and G.~Rajasekaran, \emph{{Observing the Dark Scalar Doublet
  and its Impact on the Standard-Model Higgs Boson at Colliders}},
  \href{https://doi.org/10.1103/PhysRevD.76.095011}{\emph{Phys. Rev.}
  {\bfseries D76} (2007) 095011}
  [\href{https://arxiv.org/abs/0708.2939}{{\ttfamily 0708.2939}}].

\bibitem{Bechtle:2008jh}
P.~Bechtle, O.~Brein, S.~Heinemeyer, G.~Weiglein and K.~E. Williams,
  \emph{{HiggsBounds: Confronting Arbitrary Higgs Sectors with Exclusion Bounds
  from LEP and the Tevatron}},
  \href{https://doi.org/10.1016/j.cpc.2009.09.003}{\emph{Comput.Phys.Commun.}
  {\bfseries 181} (2010) 138}
  [\href{https://arxiv.org/abs/0811.4169}{{\ttfamily 0811.4169}}].

\bibitem{Bechtle:2011sb}
P.~Bechtle, O.~Brein, S.~Heinemeyer, G.~Weiglein and K.~E. Williams,
  \emph{{HiggsBounds 2.0.0: Confronting Neutral and Charged Higgs Sector
  Predictions with Exclusion Bounds from LEP and the Tevatron}},
  \href{https://doi.org/10.1016/j.cpc.2011.07.015}{\emph{Comput.Phys.Commun.}
  {\bfseries 182} (2011) 2605}
  [\href{https://arxiv.org/abs/1102.1898}{{\ttfamily 1102.1898}}].

\bibitem{Bechtle:2013gu}
P.~Bechtle, O.~Brein, S.~Heinemeyer, O.~St{\aa}l, T.~Stefaniak, G.~Weiglein
  et~al., \emph{{Recent Developments in HiggsBounds and a Preview of
  HiggsSignals}}, \href{https://doi.org/10.22323/1.156.0024}{\emph{PoS}
  {\bfseries CHARGED2012} (2012) 024}
  [\href{https://arxiv.org/abs/1301.2345}{{\ttfamily 1301.2345}}].

\bibitem{Bechtle:2013wla}
P.~Bechtle, O.~Brein, S.~Heinemeyer, O.~Stal, T.~Stefaniak et~al.,
  \emph{{$\mathsf{HiggsBounds}-4$: Improved Tests of Extended Higgs Sectors
  against Exclusion Bounds from LEP, the Tevatron and the LHC}},
  \href{https://doi.org/10.1140/epjc/s10052-013-2693-2}{\emph{Eur.Phys.J.}
  {\bfseries C74} (2014) 2693}
  [\href{https://arxiv.org/abs/1311.0055}{{\ttfamily 1311.0055}}].

\bibitem{Bechtle:2015pma}
P.~Bechtle, S.~Heinemeyer, O.~Stal, T.~Stefaniak and G.~Weiglein,
  \emph{{Applying Exclusion Likelihoods from LHC Searches to Extended Higgs
  Sectors}}, \href{https://doi.org/10.1140/epjc/s10052-015-3650-z}{\emph{Eur.
  Phys. J.} {\bfseries C75} (2015) 421}
  [\href{https://arxiv.org/abs/1507.06706}{{\ttfamily 1507.06706}}].

\bibitem{Stal:2013hwa}
O.~St{\aa}l and T.~Stefaniak, \emph{{Constraining extended Higgs sectors with
  HiggsSignals}}, \href{https://doi.org/10.22323/1.180.0314}{\emph{PoS}
  {\bfseries EPS-HEP2013} (2013) 314}
  [\href{https://arxiv.org/abs/1310.4039}{{\ttfamily 1310.4039}}].

\bibitem{Bechtle:2013xfa}
P.~Bechtle, S.~Heinemeyer, O.~Stal, T.~Stefaniak and G.~Weiglein,
  \emph{{$HiggsSignals$: Confronting arbitrary Higgs sectors with measurements
  at the Tevatron and the LHC}},
  \href{https://doi.org/10.1140/epjc/s10052-013-2711-4}{\emph{Eur.Phys.J.}
  {\bfseries C74} (2014) 2711}
  [\href{https://arxiv.org/abs/1305.1933}{{\ttfamily 1305.1933}}].

\bibitem{Bechtle:2014ewa}
P.~Bechtle, S.~Heinemeyer, O.~St{\aa}l, T.~Stefaniak and G.~Weiglein,
  \emph{{Probing the Standard Model with Higgs signal rates from the Tevatron,
  the LHC and a future ILC}},
  \href{https://doi.org/10.1007/JHEP11(2014)039}{\emph{JHEP} {\bfseries 11}
  (2014) 039} [\href{https://arxiv.org/abs/1403.1582}{{\ttfamily 1403.1582}}].

\bibitem{Robens:2015gla}
T.~Robens and T.~Stefaniak, \emph{{Status of the Higgs Singlet Extension of the
  Standard Model after LHC Run 1}},
  \href{https://doi.org/10.1140/epjc/s10052-015-3323-y}{\emph{Eur. Phys. J.}
  {\bfseries C75} (2015) 104}
  [\href{https://arxiv.org/abs/1501.02234}{{\ttfamily 1501.02234}}].

\bibitem{Robens:2016xkb}
T.~Robens and T.~Stefaniak, \emph{{LHC Benchmark Scenarios for the Real Higgs
  Singlet Extension of the Standard Model}},
  \href{https://doi.org/10.1140/epjc/s10052-016-4115-8}{\emph{Eur. Phys. J.}
  {\bfseries C76} (2016) 268}
  [\href{https://arxiv.org/abs/1601.07880}{{\ttfamily 1601.07880}}].

\bibitem{Ilnicka:2018def}
A.~Ilnicka, T.~Robens and T.~Stefaniak, \emph{{Constraining Extended Scalar
  Sectors at the LHC and beyond}},
  \href{https://doi.org/10.1142/S0217732318300070}{\emph{Mod. Phys. Lett.}
  {\bfseries A33} (2018) 1830007}
  [\href{https://arxiv.org/abs/1803.03594}{{\ttfamily 1803.03594}}].

\bibitem{Lopez-Val:2014jva}
D.~Lopez-Val and T.~Robens, \emph{{$\Delta\,r$ and the W-boson mass in the
  singlet extension of the standard model}},
  \href{https://doi.org/10.1103/PhysRevD.90.114018}{\emph{Phys. Rev.}
  {\bfseries D90} (2014) 114018}
  [\href{https://arxiv.org/abs/1406.1043}{{\ttfamily 1406.1043}}].

\bibitem{CMS-PAS-HIG-12-045}
{\scshape CMS Collaboration} collaboration, \emph{{Combination of standard
  model Higgs boson searches and measurements of the properties of the new
  boson with a mass near 125 GeV}},  Tech. Rep. CMS-PAS-HIG-12-045, CERN,
  Geneva, 2012.

\bibitem{CMS-PAS-HIG-13-003}
{\scshape CMS Collaboration} collaboration, \emph{{Evidence for a particle
  decaying to W+W- in the fully leptonic final state in a standard model Higgs
  boson search in pp collisions at the LHC}},  Tech. Rep. CMS-PAS-HIG-13-003,
  CERN, Geneva, 2013.

\bibitem{Khachatryan:2015cwa}
{\scshape CMS} collaboration, \emph{{Search for a Higgs boson in the mass range
  from 145 to 1000 GeV decaying to a pair of W or Z bosons}},
  \href{https://doi.org/10.1007/JHEP10(2015)144}{\emph{JHEP} {\bfseries 10}
  (2015) 144} [\href{https://arxiv.org/abs/1504.00936}{{\ttfamily
  1504.00936}}].

\bibitem{Aaboud:2017rel}
{\scshape ATLAS} collaboration, \emph{{Search for heavy ZZ resonances in the
  $\ell ^+\ell ^-\ell ^+\ell ^-$ and $\ell ^+\ell ^-\nu \bar{\nu }$ final
  states using proton-proton collisions at $\sqrt{s}= 13$ $\text {TeV}$ with
  the ATLAS detector}},
  \href{https://doi.org/10.1140/epjc/s10052-018-5686-3}{\emph{Eur. Phys. J.}
  {\bfseries C78} (2018) 293}
  [\href{https://arxiv.org/abs/1712.06386}{{\ttfamily 1712.06386}}].

\bibitem{Sirunyan:2018qlb}
{\scshape CMS} collaboration, \emph{{Search for a new scalar resonance decaying
  to a pair of Z bosons in proton-proton collisions at $\sqrt{s}=13 $ TeV}},
  \href{https://doi.org/10.1007/JHEP06(2018)127,
  10.1007/JHEP03(2019)128}{\emph{JHEP} {\bfseries 06} (2018) 127}
  [\href{https://arxiv.org/abs/1804.01939}{{\ttfamily 1804.01939}}].

\bibitem{Aaboud:2018bun}
{\scshape ATLAS} collaboration, \emph{{Combination of searches for heavy
  resonances decaying into bosonic and leptonic final states using 36 fb$^{-1}$
  of proton-proton collision data at $\sqrt{s} = 13$ TeV with the ATLAS
  detector}}, \href{https://doi.org/10.1103/PhysRevD.98.052008}{\emph{Phys.
  Rev.} {\bfseries D98} (2018) 052008}
  [\href{https://arxiv.org/abs/1808.02380}{{\ttfamily 1808.02380}}].

\bibitem{Sirunyan:2018two}
{\scshape CMS} collaboration, \emph{{Combination of searches for Higgs boson
  pair production in proton-proton collisions at $\sqrt{s} = $ 13 TeV}},
  \href{https://doi.org/10.1103/PhysRevLett.122.121803}{\emph{Phys. Rev. Lett.}
  {\bfseries 122} (2019) 121803}
  [\href{https://arxiv.org/abs/1811.09689}{{\ttfamily 1811.09689}}].

\bibitem{Aad:2019uzh}
{\scshape ATLAS} collaboration, \emph{{Combination of searches for Higgs boson
  pairs in $pp$ collisions at $\sqrt{s} = $13 TeV with the ATLAS detector}},
  \href{https://arxiv.org/abs/1906.02025}{{\ttfamily 1906.02025}}.

\bibitem{z2bms}
A.~Papaefsthathiou, T.~Robens, T.~Stefaniak and J.~Zurita, ``{\sl WG
  benchmarks, $Z_2$ symmetric model}.'' Note submitted to the Higgs Cross
  Section Working Group, June 2019.

\bibitem{Ilnicka:2015jba}
A.~Ilnicka, M.~Krawczyk and T.~Robens, \emph{{Inert Doublet Model in light of
  LHC Run I and astrophysical data}},
  \href{https://doi.org/10.1103/PhysRevD.93.055026}{\emph{Phys. Rev.}
  {\bfseries D93} (2016) 055026}
  [\href{https://arxiv.org/abs/1508.01671}{{\ttfamily 1508.01671}}].

\bibitem{Abe:2018bpo}
{\scshape LHC Dark Matter Working Group} collaboration, \emph{{LHC Dark Matter
  Working Group: Next-generation spin-0 dark matter models}},
  \href{https://doi.org/10.1016/j.dark.2019.100351}{\emph{Phys. Dark Univ.}
  (2018) 100351} [\href{https://arxiv.org/abs/1810.09420}{{\ttfamily
  1810.09420}}].

\bibitem{Kalinowski:2018ylg}
J.~Kalinowski, W.~Kotlarski, T.~Robens, D.~Sokolowska and A.~F. Zarnecki,
  \emph{{Benchmarking the Inert Doublet Model for $e^+ e^-$ colliders}},
  \href{https://doi.org/10.1007/JHEP12(2018)081}{\emph{JHEP} {\bfseries 12}
  (2018) 081} [\href{https://arxiv.org/abs/1809.07712}{{\ttfamily
  1809.07712}}].

\bibitem{Dercks:2018wch}
D.~Dercks and T.~Robens, \emph{{Constraining the Inert Doublet Model using
  Vector Boson Fusion}},  \href{https://arxiv.org/abs/1812.07913}{{\ttfamily
  1812.07913}}.

\bibitem{Dolle:2009ft}
E.~Dolle, X.~Miao, S.~Su and B.~Thomas, \emph{{Dilepton Signals in the Inert
  Doublet Model}},
  \href{https://doi.org/10.1103/PhysRevD.81.035003}{\emph{Phys.Rev.} {\bfseries
  D81} (2010) 035003} [\href{https://arxiv.org/abs/0909.3094}{{\ttfamily
  0909.3094}}].

\bibitem{Swiezewska:2012eh}
B.~Swiezewska and M.~Krawczyk, \emph{{Diphoton rate in the inert doublet model
  with a 125 GeV Higgs boson}},
  \href{https://doi.org/10.1103/PhysRevD.88.035019}{\emph{Phys.Rev.} {\bfseries
  D88} (2013) 035019} [\href{https://arxiv.org/abs/1212.4100}{{\ttfamily
  1212.4100}}].

\bibitem{Gustafsson:2012aj}
M.~Gustafsson, S.~Rydbeck, L.~Lopez-Honorez and E.~Lundstrom, \emph{{Status of
  the Inert Doublet Model and the Role of multileptons at the LHC}},
  \href{https://doi.org/10.1103/PhysRevD.86.075019}{\emph{Phys. Rev.}
  {\bfseries D86} (2012) 075019}
  [\href{https://arxiv.org/abs/1206.6316}{{\ttfamily 1206.6316}}].

\bibitem{Arhrib:2012ia}
A.~Arhrib, R.~Benbrik and N.~Gaur, \emph{{$H\to \gamma \gamma$ in Inert Higgs
  Doublet Model}},
  \href{https://doi.org/10.1103/PhysRevD.85.095021}{\emph{Phys.Rev.} {\bfseries
  D85} (2012) 095021} [\href{https://arxiv.org/abs/1201.2644}{{\ttfamily
  1201.2644}}].

\bibitem{Krawczyk:2013jta}
M.~Krawczyk, D.~Sokolowska, P.~Swaczyna and B.~Swiezewska, \emph{{Constraining
  Inert Dark Matter by $R_{\gamma\gamma}$ and WMAP data}},
  \href{https://doi.org/10.1007/JHEP09(2013)055}{\emph{JHEP} {\bfseries 1309}
  (2013) 055} [\href{https://arxiv.org/abs/1305.6266}{{\ttfamily 1305.6266}}].

\bibitem{Belanger:2015kga}
G.~Belanger, B.~Dumont, A.~Goudelis, B.~Herrmann, S.~Kraml and D.~Sengupta,
  \emph{{Dilepton constraints in the Inert Doublet Model from Run 1 of the
  LHC}}, \href{https://doi.org/10.1103/PhysRevD.91.115011}{\emph{Phys. Rev.}
  {\bfseries D91} (2015) 115011}
  [\href{https://arxiv.org/abs/1503.07367}{{\ttfamily 1503.07367}}].

\bibitem{deFlorian:2016spz}
{\scshape LHC Higgs Cross Section Working Group} collaboration, \emph{{Handbook
  of LHC Higgs Cross Sections: 4. Deciphering the Nature of the Higgs Sector}},
   \href{https://arxiv.org/abs/1610.07922}{{\ttfamily 1610.07922}}.

\bibitem{Poulose:2016lvz}
P.~Poulose, S.~Sahoo and K.~Sridhar, \emph{{Exploring the Inert Doublet Model
  through the dijet plus missing transverse energy channel at the LHC}},
  \href{https://doi.org/10.1016/j.physletb.2016.12.022}{\emph{Phys. Lett.}
  {\bfseries B765} (2017) 300}
  [\href{https://arxiv.org/abs/1604.03045}{{\ttfamily 1604.03045}}].

\bibitem{Datta:2016nfz}
A.~Datta, N.~Ganguly, N.~Khan and S.~Rakshit, \emph{{Exploring collider
  signatures of the inert Higgs doublet model}},
  \href{https://doi.org/10.1103/PhysRevD.95.015017}{\emph{Phys. Rev.}
  {\bfseries D95} (2017) 015017}
  [\href{https://arxiv.org/abs/1610.00648}{{\ttfamily 1610.00648}}].

\bibitem{Kanemura:2016sos}
S.~Kanemura, M.~Kikuchi and K.~Sakurai, \emph{{Testing the dark matter scenario
  in the inert doublet model by future precision measurements of the Higgs
  boson couplings}},
  \href{https://doi.org/10.1103/PhysRevD.94.115011}{\emph{Phys. Rev.}
  {\bfseries D94} (2016) 115011}
  [\href{https://arxiv.org/abs/1605.08520}{{\ttfamily 1605.08520}}].

\bibitem{Akeroyd:2016ymd}
A.~G. Akeroyd et~al., \emph{{Prospects for charged Higgs searches at the LHC}},
  \href{https://doi.org/10.1140/epjc/s10052-017-4829-2}{\emph{Eur. Phys. J.}
  {\bfseries C77} (2017) 276}
  [\href{https://arxiv.org/abs/1607.01320}{{\ttfamily 1607.01320}}].

\bibitem{Wan:2018eaz}
N.~Wan, N.~Li, B.~Zhang, H.~Yang, M.-F. Zhao, M.~Song et~al., \emph{{Searches
  for Dark Matter via Mono-W Production in Inert Doublet Model at the LHC}},
  \href{https://doi.org/10.1088/0253-6102/69/5/617}{\emph{Commun. Theor. Phys.}
  {\bfseries 69} (2018) 617}.

\bibitem{Belyaev:2018ext}
A.~Belyaev, T.~R. Fernandez Perez~Tomei, P.~G. Mercadante, C.~S. Moon,
  S.~Moretti, S.~F. Novaes et~al., \emph{{Advancing LHC probes of dark matter
  from the inert two-Higgs-doublet model with the monojet signal}},
  \href{https://doi.org/10.1103/PhysRevD.99.015011}{\emph{Phys. Rev.}
  {\bfseries D99} (2019) 015011}
  [\href{https://arxiv.org/abs/1809.00933}{{\ttfamily 1809.00933}}].

\bibitem{Bhardwaj:2019mts}
A.~Bhardwaj, P.~Konar, T.~Mandal and S.~Sadhukhan, \emph{{Probing Inert Doublet
  Model using jet substructure with multivariate analysis}},
  \href{https://arxiv.org/abs/1905.04195}{{\ttfamily 1905.04195}}.

\bibitem{Altarelli:1990zd}
G.~Altarelli and R.~Barbieri, \emph{{Vacuum polarization effects of new physics
  on electroweak processes}},
  \href{https://doi.org/10.1016/0370-2693(91)91378-9}{\emph{Phys.~Lett.~B}
  {\bfseries 253} (1991) 161}.

\bibitem{Peskin:1990zt}
M.~E. Peskin and T.~Takeuchi, \emph{{A New constraint on a strongly interacting
  Higgs sector}},
  \href{https://doi.org/10.1103/PhysRevLett.65.964}{\emph{Phys.Rev.Lett.}
  {\bfseries 65} (1990) 964}.

\bibitem{Peskin:1991sw}
M.~E. Peskin and T.~Takeuchi, \emph{{Estimation of oblique electroweak
  corrections}},
  \href{https://doi.org/10.1103/PhysRevD.46.381}{\emph{Phys.Rev.} {\bfseries
  D46} (1992) 381}.

\bibitem{Maksymyk:1993zm}
I.~Maksymyk, C.~Burgess and D.~London, \emph{{Beyond S, T and U}},
  \href{https://doi.org/10.1103/PhysRevD.50.529}{\emph{Phys.Rev.} {\bfseries
  D50} (1994) 529} [\href{https://arxiv.org/abs/hep-ph/9306267}{{\ttfamily
  hep-ph/9306267}}].

\bibitem{Eriksson:2009ws}
D.~Eriksson, J.~Rathsman and O.~Stal, \emph{{2HDMC: Two-Higgs-Doublet Model
  Calculator Physics and Manual}},
  \href{https://doi.org/10.1016/j.cpc.2009.09.011}{\emph{Comput.Phys.Commun.}
  {\bfseries 181} (2010) 189}
  [\href{https://arxiv.org/abs/0902.0851}{{\ttfamily 0902.0851}}].

\bibitem{Barducci:2016pcb}
D.~Barducci, G.~Belanger, J.~Bernon, F.~Boudjema, J.~Da~Silva, S.~Kraml et~al.,
  \emph{{Collider limits on new physics within micrOMEGAs$\_$4.3}},
  \href{https://doi.org/10.1016/j.cpc.2017.08.028}{\emph{Comput. Phys. Commun.}
  {\bfseries 222} (2018) 327}
  [\href{https://arxiv.org/abs/1606.03834}{{\ttfamily 1606.03834}}].

\bibitem{trtalk}
T.~Robens, ``{{\sl IDM benchmarks for the LHC at 13 and 27 TeV}, Talk at the
  Higgs Cross Section working group WG3 subgroup meeting, 24.10.18}.''

\bibitem{Sirunyan:2018owy}
{\scshape CMS} collaboration, \emph{{Search for invisible decays of a Higgs
  boson produced through vector boson fusion in proton-proton collisions at
  $\sqrt{s} =$ 13 TeV}},
  \href{https://doi.org/10.1016/j.physletb.2019.04.025}{\emph{Phys. Lett.}
  {\bfseries B793} (2019) 520}
  [\href{https://arxiv.org/abs/1809.05937}{{\ttfamily 1809.05937}}].

\bibitem{ATLAS-CONF-2017-060}
{\scshape ATLAS Collaboration} collaboration, \emph{{Search for dark matter and
  other new phenomena in events with an energetic jet and large missing
  transverse momentum using the ATLAS detector}},  Tech. Rep.
  ATLAS-CONF-2017-060, CERN, Geneva, Jul, 2017.

\bibitem{Aaboud:2017phn}
{\scshape ATLAS} collaboration, \emph{{Search for dark matter and other new
  phenomena in events with an energetic jet and large missing transverse
  momentum using the ATLAS detector}},
  \href{https://doi.org/10.1007/JHEP01(2018)126}{\emph{JHEP} {\bfseries 01}
  (2018) 126} [\href{https://arxiv.org/abs/1711.03301}{{\ttfamily
  1711.03301}}].

\bibitem{Kalinowski:2018kdn}
J.~Kalinowski, W.~Kotlarski, T.~Robens, D.~Sokolowska and A.~F. Zarnecki,
  \emph{{Exploring Inert Scalars at CLIC}},
  \href{https://doi.org/10.1007/JHEP07(2019)053}{\emph{JHEP} {\bfseries 07}
  (2019) 053} [\href{https://arxiv.org/abs/1811.06952}{{\ttfamily
  1811.06952}}].

\bibitem{deBlas:2018mhx}
J.~de~Blas et~al., \emph{{The CLIC Potential for New Physics}},
  \href{https://arxiv.org/abs/1812.02093}{{\ttfamily 1812.02093}}.

\bibitem{Zarnecki:2019poj}
A.~F. Zarnecki, J.~Kalinowski, J.~Klamka, P.~Sopicki, W.~Kotlarski, T.~Robens
  et~al., \emph{{Inert Doublet Model Signatures at Future e+e- Colliders}},  in
  \emph{{An Alpine LHC Physics Summit 2019 (ALPS 2019) Obergurgl, Austria,
  April 22-27, 2019,{PoS(ALPS2019)010}}}, 2019,
  \href{https://arxiv.org/abs/1908.04659}{{\ttfamily 1908.04659}}.

\bibitem{dorotatalk}
D.~Sokolowska, ``{{\sl Inert Doublet Signatures at Future $e^+e^-$ Colliders},
  Talk at the EPS-HEP 2019 conference, 12.7.19, to be published in proceedings
  of the European Physical Society Conference on High Energy Physics 2019
  (EPS-HEP2019) {PoS(EPS-HEP2019)570} (2019)}.''

\bibitem{Staub:2015kfa}
F.~Staub, \emph{{Exploring new models in all detail with SARAH}},
  \href{https://doi.org/10.1155/2015/840780}{\emph{Adv. High Energy Phys.}
  {\bfseries 2015} (2015) 840780}
  [\href{https://arxiv.org/abs/arXiv:1503.04200}{{\ttfamily
  arXiv:1503.04200}}].

\bibitem{Porod:2003um}
W.~Porod, \emph{{SPheno, a program for calculating supersymmetric spectra, SUSY
  particle decays and SUSY particle production at e+ e- colliders}},
  \href{https://doi.org/10.1016/S0010-4655(03)00222-4}{\emph{Comput. Phys.
  Commun.} {\bfseries 153} (2003) 275}
  [\href{https://arxiv.org/abs/hep-ph/0301101}{{\ttfamily hep-ph/0301101}}].

\bibitem{Porod:2011nf}
W.~Porod and F.~Staub, \emph{{SPheno 3.1: Extensions including flavour,
  CP-phases and models beyond the MSSM}},
  \href{https://doi.org/10.1016/j.cpc.2012.05.021}{\emph{Comput. Phys. Commun.}
  {\bfseries 183} (2012) 2458}
  [\href{https://arxiv.org/abs/1104.1573}{{\ttfamily 1104.1573}}].

\bibitem{Moretti:2001zz}
M.~Moretti, T.~Ohl and J.~Reuter, \emph{{O'Mega: An Optimizing matrix element
  generator}},  2001.

\bibitem{Kilian:2007gr}
W.~Kilian, T.~Ohl and J.~Reuter, \emph{{WHIZARD: Simulating Multi-Particle
  Processes at LHC and ILC}},
  \href{https://doi.org/10.1140/epjc/s10052-011-1742-y}{\emph{Eur. Phys. J.}
  {\bfseries C71} (2011) 1742}
  [\href{https://arxiv.org/abs/0708.4233}{{\ttfamily 0708.4233}}].

\bibitem{Hocker:2007ht}
A.~Hocker et~al., \emph{{TMVA - Toolkit for Multivariate Data Analysis}},
  \href{https://arxiv.org/abs/physics/0703039}{{\ttfamily physics/0703039}}.

\bibitem{Aaboud:2018iil}
{\scshape ATLAS} collaboration, \emph{{Search for the Higgs boson produced in
  association with a vector boson and decaying into two spin-zero particles in
  the $H \rightarrow aa \rightarrow 4b$ channel in $pp$ collisions at $\sqrt{s}
  = 13$ TeV with the ATLAS detector}},
  \href{https://doi.org/10.1007/JHEP10(2018)031}{\emph{JHEP} {\bfseries 10}
  (2018) 031} [\href{https://arxiv.org/abs/1806.07355}{{\ttfamily
  1806.07355}}].

\bibitem{Aaboud:2018esj}
{\scshape ATLAS} collaboration, \emph{{Search for Higgs boson decays into a
  pair of light bosons in the $bb\mu\mu$ final state in $pp$ collision at
  $\sqrt{s} = $13 TeV with the ATLAS detector}},
  \href{https://doi.org/10.1016/j.physletb.2018.10.073}{\emph{Phys. Lett.}
  {\bfseries B790} (2019) 1}
  [\href{https://arxiv.org/abs/1807.00539}{{\ttfamily 1807.00539}}].

\bibitem{Sirunyan:2018mot}
{\scshape CMS} collaboration, \emph{{Search for an exotic decay of the Higgs
  boson to a pair of light pseudoscalars in the final state with two muons and
  two b quarks in pp collisions at 13 TeV}},
  \href{https://doi.org/10.1016/j.physletb.2019.06.021}{\emph{Phys. Lett.}
  {\bfseries B795} (2019) 398}
  [\href{https://arxiv.org/abs/1812.06359}{{\ttfamily 1812.06359}}].

\bibitem{Sirunyan:2019gou}
{\scshape CMS} collaboration, \emph{{Search for light pseudoscalar boson pairs
  produced from decays of the 125 GeV Higgs boson in final states with two
  muons and two nearby tracks in pp collisions at $\sqrt{s}=$ 13 TeV}},
  \href{https://arxiv.org/abs/1907.07235}{{\ttfamily 1907.07235}}.

\bibitem{Barger:2008jx}
V.~Barger, P.~Langacker, M.~McCaskey, M.~Ramsey-Musolf and G.~Shaughnessy,
  \emph{{Complex Singlet Extension of the Standard Model}},
  \href{https://doi.org/10.1103/PhysRevD.79.015018}{\emph{Phys. Rev.}
  {\bfseries D79} (2009) 015018}
  [\href{https://arxiv.org/abs/0811.0393}{{\ttfamily 0811.0393}}].

\bibitem{Coimbra:2013qq}
R.~Coimbra, M.~O.~P. Sampaio and R.~Santos, \emph{{ScannerS: Constraining the
  phase diagram of a complex scalar singlet at the LHC}},
  \href{https://doi.org/10.1140/epjc/s10052-013-2428-4}{\emph{Eur. Phys. J.}
  {\bfseries C73} (2013) 2428}
  [\href{https://arxiv.org/abs/1301.2599}{{\ttfamily 1301.2599}}].

\bibitem{Costa:2015llh}
R.~Costa, M.~M\"uhlleitner, M.~O.~P. Sampaio and R.~Santos, \emph{{Singlet
  Extensions of the Standard Model at LHC Run 2: Benchmarks and Comparison with
  the NMSSM}}, \href{https://doi.org/10.1007/JHEP06(2016)034}{\emph{JHEP}
  {\bfseries 06} (2016) 034}
  [\href{https://arxiv.org/abs/1512.05355}{{\ttfamily 1512.05355}}].

\bibitem{Robens:2019kga}
T.~Robens, T.~Stefaniak and J.~Wittbrodt, \emph{{Two-real-scalar-singlet
  extension of the SM: LHC phenomenology and benchmark scenarios}},
  \href{https://arxiv.org/abs/1908.08554}{{\ttfamily 1908.08554}}.

\bibitem{Ferreira:2014dya}
P.~M. Ferreira, R.~Guedes, M.~O.~P. Sampaio and R.~Santos, \emph{{Wrong sign
  and symmetric limits and non-decoupling in 2HDMs}},
  \href{https://doi.org/10.1007/JHEP12(2014)067}{\emph{JHEP} {\bfseries 12}
  (2014) 067} [\href{https://arxiv.org/abs/1409.6723}{{\ttfamily 1409.6723}}].

\bibitem{Muhlleitner:2016mzt}
M.~M{\"{u}}hlleitner, M.~O.~P. Sampaio, R.~Santos and J.~Wittbrodt, \emph{{The
  N2HDM under Theoretical and Experimental Scrutiny}},
  \href{https://doi.org/10.1007/JHEP03(2017)094}{\emph{JHEP} {\bfseries 03}
  (2017) 094} [\href{https://arxiv.org/abs/1612.01309}{{\ttfamily
  1612.01309}}].

\bibitem{Aaboud:2018ksn}
{\scshape ATLAS} collaboration, \emph{{Search for Higgs boson pair production
  in the $WW^{(*)}WW^{(*)}$ decay channel using ATLAS data recorded at
  $\sqrt{s}=13$ TeV}},
  \href{https://doi.org/10.1007/JHEP05(2019)124}{\emph{JHEP} {\bfseries 05}
  (2019) 124} [\href{https://arxiv.org/abs/1811.11028}{{\ttfamily
  1811.11028}}].

\end{thebibliography}
\end{document}